\documentclass[aps,twocolumn,showpacs,preprintnumbers,floats]{revtex4}\def\@cite#1#2{\textsuperscript{[{#1\if@tempswa , #2\fi}]}}

\usepackage{graphicx}
\usepackage{amsmath}
\usepackage{amsfonts}
\usepackage{amssymb}
\usepackage{color}
\usepackage{subfigure}
\usepackage{epsfig}
\usepackage{morefloats}
\usepackage{multirow}
\usepackage{graphicx,booktabs}
\usepackage{mathrsfs}
\usepackage{txfonts}
\usepackage{indentfirst}
\usepackage{graphicx,booktabs}
\usepackage[colorlinks, citecolor=blue,anchorcolor=red,menucolor=red, linkcolor=red,filecolor=red,urlcolor=blue,frenchlinks=red]{hyperref}
\usepackage{longtable,lscape}
\usepackage{warpcol}
\usepackage{dcolumn}
\usepackage{longtable}

\newcommand{\vsig}{\mbox{\boldmath$\sigma$\unboldmath}}
\newcommand{\veps}{\mbox{\boldmath$\epsilon$\unboldmath}}

\begin{document}

\title{Towards establishing an abundant $B$ and $B_s$ spectrum up to the second orbital excitations}
\author{Qi Li, Ru-Hui Ni, Xian-Hui Zhong \footnote {mail: zhongxh@hunnu.edu.cn}
 }  \affiliation{ 1) Department
of Physics, Hunan Normal University,  Changsha 410081, China }

\affiliation{ 2) Synergetic Innovation
Center for Quantum Effects and Applications (SICQEA), Changsha 410081,China}

\affiliation{ 3) Key Laboratory of
Low-Dimensional Quantum Structures and Quantum Control of Ministry
of Education, Changsha 410081, China}

%\date{\today}

\begin{abstract}

Stimulated by the exciting progress in experiments, we carry out a combined analysis of the masses, and strong and radiative decay properties of the $B$ and $B_s$-meson states up to the second orbital excitations. Based on our good descriptions of the mass
and decay properties for the low-lying well-established states $B_1(5721)$, $B_2^*(5747)$, $B_{s1}(5830)$ and $B_{s2}^*(5840)$,
we give a quark model classification for the high mass resonances observed in recent years. It is found that (i) the $B_{J}(5840)$ resonance may be explained as the low mass mixed state $B(|SD\rangle_L)$ via $2^3S_1$-$1^3D_1$ mixing, or the pure $B(2^3S_1)$ state, or $B(2^1S_0)$. (ii) The $B_J(5970)$ resonance may be assigned as the $1^3D_3$ state in the $B$ meson family, although it as a pure $2^3S_1$ state cannot be excluded. (iii) The narrow structure around 6064 MeV observed in the $B^+K^-$ mass spectrum at LHCb may be mainly caused by the $B_{sJ}(6109)$ resonance decaying into $B^{*+}K^-$, and favors the assignment of the high mass $1D$-wave mixed state $B_s(1D'_2)$ with $J^P=2^-$, although it as the $1^3D_3$ state cannot be excluded. (iv) The relatively broader $B_{sJ}(6114)$ structure observed at LHCb may be explained with the mixed state $B_s(|SD\rangle_H)$ via $2^3S_1$-$1^3D_1$ mixing, or a pure $1^3D_1$ state. Most of the missing $1P$-, $1D$-, and $2S$-wave $B$- and $B_s$-meson states have a relatively narrow width, they are most likely to be observed in their dominant decay channels with a larger data sample at LHCb.

\end{abstract}

\maketitle

\section{Introduction}

Since 2007, significant progress has been made in the observations of the bottom and
bottom-strange mesons~\cite{Zyla:2020zbs}. In 2007, two low-lying orbitally excited narrow $B$
mesons $B_1(5721)^{0,+}$ and $B_2^*(5747)^{0,+}$ were observed
by the D0 experiment~\cite{Abazov:2007vq}, and were confirmed by the CDF
experiment one year later~\cite{Aaltonen:2008aa}. Their strange analogues, $B_{s1}(5830)$
and $B_{s2}^*(5840)$, as the first orbitally excited $B_s$ mesons, were also
reported by the CDF Collaboration in 2007~\cite{Aaltonen:2007ah}.
The $B_{s1}(5830)$ and $B_{s2}^*(5840)$ were confirmed by the D0 and LHCb experiments~\cite{Aaij:2012uva,Abazov:2007af}.
In 2013, two higher resonances $B(5970)^{0,+}$ were observed in the $B\pi$ final states by the
CDF Collaboration~\cite{Aaltonen:2013atp}. In 2015, four higher resonances
$B_J(5840)^{0,+}$ and $B_J(5960)^{0,+}$ were observed in the $B\pi$ final states by the
LHCb Collaboration when they carried out precise measurements of the
properties of the $B_1(5721)^{0,+}$ and $B_2^*(5747)^{0,+}$ states~\cite{Aaij:2015qla}.
The properties of the $B_J(5960)^{0,+}$ states are consistent with those of
$B(5970)^{0,+}$ obtained by the CDF Collaboration.
Recently, the LHCb Collaboration observed two structures $B_{sJ}(6064)$ and $B_{sJ}(6114)$ in
the $B^+K^-$ mass spectrum~\cite{Aaij:2020hcw}. More experimental
information about the excited bottom and bottom-strange mesons
is collected in Table~\ref{Excitation}. More and more excited bottom and bottom-strange mesons
are expected to be observed in future LHCb experiments due to its huge production cross
sections of beauty, together with a good reconstruction efficiency, versatile trigger scheme
and an excellent momentum and mass resolution~\cite{Belyaev:2021cyr}.

\begin{table*}[ht]
\caption{Summary of the experimental information for the excited $B$- and $B_s$-meson states.
The date for the $B_1(5721)^{+,0}$, $B_2(5747)^{+,0}$, $B_{s1}(5830)^0$, and $B_{s2}(5840)^0$
resonances are adopted the average values of the Review of Particle Physics (RPP) of Particle
Data Group (PDG)~\cite{Zyla:2020zbs}. The N and UN stand for the natural spin parity $P=(-1)^J$
and unnatural spin parity $P=-(-1)^J$.} \label{Excitation}
\begin{tabular}{cccccccc }\hline\hline
&~~~~~Resonance~~~~ &~~~~$J^P$~~~~ &~~~~Mass (MeV) ~~~~& ~~~~Width (MeV) ~~~~& ~~~~ Observed channel~~~~ & Experiment  \\
\hline
&$B_1(5721)^0$ &$1^+$ & $5726.1\pm1.3 $ & $27.5\pm 3.4$ &$B^{*+}\pi^-$    & D0~\cite{Abazov:2007vq},CDF~\cite{Aaltonen:2008aa},LHCb~\cite{Aaij:2015qla}   \\
&$B_1(5721)^+$ &$1^+$ &$5725.9^{+2.5}_{-2.7} $ & $31\pm 6$ &$B^{*0}\pi^+$  & D0~\cite{Abazov:2007vq},CDF~\cite{Aaltonen:2008aa},LHCb~\cite{Aaij:2015qla}     \\
&$B_2(5747)^0$ &$2^+$ &$5739.5\pm0.7 $ & $24.2\pm 1.7$ &$B^{+}\pi^-$,$B^{*+}\pi^-$ & D0~\cite{Abazov:2007vq},CDF~\cite{Aaltonen:2008aa},LHCb~\cite{Aaij:2015qla}      \\
&$B_2(5747)^+$ &$2^+$ &$5737.2\pm0.7 $ & $20\pm 5$ &$B^{+}\pi^-$,$B^{*+}\pi^-$ & D0~\cite{Abazov:2007vq},CDF~\cite{Aaltonen:2008aa},LHCb~\cite{Aaij:2015qla}       \\
%&$B_J(5732)^+$ &$?$ &$5698\pm 8 $ & $128\pm 18$ &$B^{*}\pi+B\pi$,$B^{*}\pi$ & OPAL,CDF,ALEPH,DELPHI,L3      \\

&$B_J(5970)^+$ &$?$ &$5961\pm 17 $ & $60^{+30}_{-20}\pm 40$ &$B^{0}\pi^+$ [or $B^{*0}\pi^+$]& CDF~\cite{Aaltonen:2013atp}      \\
&$B_J(5970)^0$ &$?$ &$5978\pm 17 $ & $70^{+30}_{-20}\pm 30$ &$B^{+}\pi^-$ [or $B^{*+}\pi^-$]& CDF~\cite{Aaltonen:2013atp}      \\
\hline
Case A &$B_J(5960)^+$ &UN &$5964.9\pm 6.8 $ & $63.0\pm 31.7$ &$B^{*0}\pi^+$ & LHCb  ~\cite{Aaij:2015qla}    \\
&$B_J(5960)^0$ &UN &$5969.2\pm 8.2 $ & $82.3\pm 17.1$ &$B^{*+}\pi^-$  & LHCb ~\cite{Aaij:2015qla}    \\
&$B_J(5840)^+$ &UN &$5850.3\pm 26.6 $ & $224.4\pm 103.7$ &$B^{*0}\pi^+$ & LHCb ~\cite{Aaij:2015qla}     \\
&$B_J(5840)^0$ &UN &$5862.9\pm 11.9 $ & $127.4\pm 50.9$ &$B^{*+}\pi^-$  & LHCb ~\cite{Aaij:2015qla}    \\

Case B&$ B_J(5960)^+$ &UN &$6010.6\pm 7.1 $ & $61.4\pm 31.7$ &$B^{*0}\pi^+$ & LHCb ~\cite{Aaij:2015qla}     \\
&$B_J(5960)^0$ &UN &$6015.9\pm 9.2 $ & $81.6\pm 19.3$ &$B^{*+}\pi^-$  & LHCb ~\cite{Aaij:2015qla}    \\
&$B_J(5840)^+$ &N &$5874.5\pm 39.6 $ & $214.6\pm 106.5$ &$B^{*0}\pi^+$, $B^{0}\pi^+$? & LHCb  ~\cite{Aaij:2015qla}    \\
&$B_J(5840)^0$ &N &$5889.7\pm 29.1 $ & $107.0\pm 53.8$ &$B^{*+}\pi^-$, $B^{+}\pi^-$?   & LHCb ~\cite{Aaij:2015qla}    \\

Case C&$ B_J(5960)^+$ &N &$5966.4\pm 7.2 $ & $60.8\pm 31.2$ &$B^{*0}\pi^+$, $B^{0}\pi^+$? & LHCb ~\cite{Aaij:2015qla}     \\
&$B_J(5960)^0$ &N &$5993.6\pm 11.7 $ & $55.9\pm 16.0$ &$B^{*+}\pi^-$, $B^{+}\pi^-$?  & LHCb ~\cite{Aaij:2015qla}    \\
&$B_J(5840)^+$ &UN &$5889.3\pm 29.3 $ & $229.3\pm 104.7$ &$B^{*0}\pi^+$  & LHCb  ~\cite{Aaij:2015qla}    \\
&$B_J(5840)^0$ &UN &$5907.8\pm 13.0 $ & $119.4\pm 51.4$ &$B^{*+}\pi^-$   & LHCb  ~\cite{Aaij:2015qla}   \\
\hline
&$B_{s1}(5830)^0$&$1^+$ & $5828.70\pm 0.20 $ & $0.5\pm 0.3\pm 0.3$ &$B^{*+}K^-$, $B^{*0}K^0$ & CDF~\cite{Aaltonen:2007ah}, D0 ~\cite{Abazov:2007af}, LHCb~\cite{Aaij:2012uva}    \\
&$B_{s2}(5840)^0$&$2^+$ & $5839.86\pm 0.12 $ & $1.49\pm 0.27$ &$B^{*}K$, $BK$& CDF~\cite{Aaltonen:2007ah},
D0~\cite{Abazov:2007af},LHCb~\cite{Aaij:2012uva}        \\
\hline
%$B_{sJ}(5850)^0$&$?$ & $5853\pm 15 $ & $47\pm 22$ &   & OPAL   \\
&$B_{sJ}(6064)^0$&$?$ & $6063.5\pm 2.0$ & $26\pm 4\pm 4$ & $B^{+}K^-$ & LHCb~\cite{Aaij:2020hcw}    \\
  &[or $B_{sJ}(6109)^0$]&$?$ & [or $6108.8\pm 1.8$] & [or $22\pm 5\pm 4$] &  [or $B^{*+}K^-$] & LHCb~\cite{Aaij:2020hcw}    \\
&$B_{sJ}(6114)^0$&$?$ & $6114\pm 8$ & $66\pm 18\pm 21$ & $B^{+}K^-$ & LHCb~\cite{Aaij:2020hcw}    \\
 &[or $B_{sJ}(6158)^0$]&$?$ & [or $6158\pm 9$] & [or $72\pm 18\pm 25$] &  [or $B^{*+}K^-$] & LHCb~\cite{Aaij:2020hcw}    \\
\hline\hline
\end{tabular}
\end{table*}

In theory, many theoretical studies of the masses~\cite{Godfrey:1985xj,Zeng:1994vj,DiPierro:2001dwf,Jia:2018vwl,Godfrey:2019cmi,Gregory:2010gm,Lang:2015hza,Shah:2016mgq,
Lu:2017hma,Yazarloo:2016luc,Alhakami:2020vil,Ebert:2009ua,Cheng:2017oqh,Liu:2016efm,Asghar:2018tha,Lu:2016bbk,Sun:2014wea,Godfrey:2016nwn,Kher:2017mky},
strong decays~\cite{Zhu:1998wy,Orsland:1998de,DiPierro:2001dwf,Sun:2014wea,Lu:2016bbk,Godfrey:2016nwn,
Asghar:2018tha,Gupta:2018xds,Godfrey:2019cmi,Yu:2019iwm,
Gupta:2017bcm,Zhang:2018ubo,Alhendi:2015rka,Ferretti:2015rsa,Wang:2014oca,Xu:2014mka,Wang:2014cta,Zhang:2010iea,Luo:2009wu,Yu:2019sqp}, radiative decays~\cite{Orsland:1998de,Godfrey:2019cmi,Shah:2016mgq,Lu:2016bbk,Yu:2019sqp,Aliev:2018kry,Godfrey:2016nwn,Asghar:2018tha}, and weak decays~\cite{Gan:2010hw,Shah:2016mgq,Grinstein:2015aua} for the excited bottom and bottom-strange
meson states have been carried out with different methods.
For the well established states $B_1(5721)$, $B_2^*(5747)$, $B_{s1}(5830)$ and $B_{s2}^*(5840)$, there are no puzzles
to classify them as the first orbital excitations (i.e. the $1P$-wave states) of quark models.
While for the newly observed resonances/structures $B_J(5840)^{0,+}$, $B(5970)^{0,+}$,
$B_{sJ}(6064)$, and $B_{sJ}(6114)$, although they are good candidates of the $2S$
and $1D$-wave states according to the mass spectrum predictions
in various quark models, their quark model classification is not clear. There are some theoretical interpretations
of the newly observed $B_J(5840)$ and $B_J(5970)$ based on the predicted masses and strong
decay properties, since they have been reported by the CDF and LHCb experiments.
In the literature, the $B_J(5840)$ resonance is explained with the $B(2^1S_0)$~\cite{Gupta:2017bcm,Lu:2016bbk,Asghar:2018tha},
the $B(2^3S_1)$~\cite{Yu:2019iwm}, or the $B(1^3D_1)$ state~\cite{Gupta:2018xds}. While for the resonance $B_J(5970)$, there
are interpretations with the radially excited state $B(2^3S_1)$~\cite{Sun:2014wea,Gupta:2017bcm,Liu:2016efm},
or with the second orbitally excited $B$-meson states~\cite{Godfrey:2019cmi} either $B(1^3D_3)$~\cite{Yu:2019iwm,Lu:2016bbk} or $B(1^3D_1)$~\cite{Asghar:2018tha}. It should be mentioned that in Ref.~\cite{Xiao:2014ura}, our group assigned the $B_J(5970)$ resonance
to be the $B(1^3D_3)$ state by analyzing the strong decay properties within
a chiral quark model. Based on this assignment, the authors further predicted
that as the partner of $B_J(5970)$, the mass and width of the $B_s(1^3D_3)$ state
might be $M\simeq 6.07$ GeV and $\Gamma\simeq 30$ MeV, respectively, which are also
in agreement with those predictions in Refs.~\cite{Lu:2016bbk,Asghar:2018tha}.
If assigning the $B_{sJ}(6064)$ to be the $B_s(1^3D_3)$ state,
both the measured mass and width are consistent with quark model predictions.
It should be mentioned that there are no discussions about the recently observed structures $B_{sJ}(6064)$
and $B_{sJ}(6114)$ in the literature. More information about the status of the bottom and
bottom-strange meson study can be found in the recent review work~\cite{Chen:2016spr}.

The experimental progress provides us good opportunities to establish an abundant $B$ and $B_s$-meson spectrum
up to the second orbital ($L=2$) excitations. In this work we deepen our study by
carrying out a combined analysis of the masses and decay properties
of the $B$ and $B_s$-meson states up to the $L=2$ excitations.
First, we calculate the mass spectrum of $B$ and $B_s$ mesons within a
nonrelativistic potential model. With this model the masses for the
observed $B$ and $B_s$-meson states can be described successfully.
Then, with the available wave functions from the potential model,
we calculate the OZI-allowed two-body strong decays of the excited
$B$ and $B_s$ mesons with a chiral quark model~\cite{Manohar:1983md,Li:1994cy,Li:1997gd,Zhao:2002id}.
This model has been successfully applied to describe the strong decays of the heavy-light mesons and
baryons~\cite{DiPierro:2001dwf,Xiao:2014ura,Zhong:2010vq,Zhong:2008kd,
Zhong:2009sk,Liu:2012sj,Zhong:2007gp,Xiao:2013xi,Nagahiro:2016nsx,Yao:2018jmc,
Wang:2017kfr,Xiao:2017udy,Wang:2017hej,Liu:2019wdr}.
To provide more knowledge for the excited $B$ and $B_s$ meson states,
we also evaluate the their electromagnetic (EM) transitions within a nonrelativistic
constituent quark model~\cite{Brodsky:1968ea,Close:1970kt,Close:1989aj,Li:1994cy,Li:1993zzb,Li:1997gd,
Zhao:1998fn,Zhao:2000tb,Zhao:2001jw,Zhao:2002id}. This model has also
been successfully applied to describe the radiative decays of baryon
states~\cite{Yao:2018jmc,Wang:2017kfr,Lu:2017meb,Xiao:2017udy,
Wang:2017hej,Liu:2019vtx,Liu:2019wdr} and meson systems~\cite{Deng:2016stx,
Deng:2016ktl,Li:2019tbn,Close:2002ky,Chen:2020jku}. Based on our good descriptions of the mass
and decay properties for the low-lying well-established states
$B_1(5721)^{0,+}$, $B_2^*(5747)^{0,+}$, $B_{s1}(5830)$ and $B_{s2}^*(5840)$,
we give our quark model classifications of the high mass resonances/structures
$B_J(5840)^{0,+}$, $B(5970)^{0,+}$, $B_{sJ}(6064)$, and $B_{sJ}(6114)$.
Finally, according to our assignments for the newly observed resonances,
we attempt to predict the properties of the missing resonances, which may
be useful for future investigations in experiments.

This paper is organized as follows. In Sec.~\ref{spectrum}, the mass spectrum is calculated within
a non-relativistic linear potential model. In Sec.~\ref{decay}, a brief review of the
chiral quark model is given. The numerical results are presented and discussed in Sec.~\ref{DIS}.
Finally, a summary is given in Sec.~\ref{SUM}.

\begin{table}[htb]
\begin{center}
\caption{The parameters of the nonrelativistic potential model.}\label{parameter}
\begin{tabular}{ccccccccccccccc}
\hline\hline
~~~~~~~~~~~~~~~~~~~&$B$~~~~~~~~~~~~~~~~~~~& $B_s$  \\
\hline
$m_b$ (GeV)            ~~~~~~~~~~~~~~~~~~~&   4.852         ~~~~~~~~~~~~~~~~~~& 4.852       \\
$m_{u,d}$ (GeV)        ~~~~~~~~~~~~~~~~~~~&   0.450         ~~~~~~~~~~~~~~~~~~& $\cdots$     \\
$m_s$ (GeV)            ~~~~~~~~~~~~~~~~~~~&   $\cdots$       ~~~~~~~~~~~~~~~~~~& 0.600       \\
$\alpha$               ~~~~~~~~~~~~~~~~~~~&   0.564         ~~~~~~~~~~~~~~~~~~& 0.550       \\
$\sigma$ (GeV)         ~~~~~~~~~~~~~~~~~~~&   0.98          ~~~~~~~~~~~~~~~~~~& 1.06        \\
$b$      (GeV$^2$)     ~~~~~~~~~~~~~~~~~~~&   0.120         ~~~~~~~~~~~~~~~~~~& 0.120       \\
$C_0$    (GeV)         ~~~~~~~~~~~~~~~~~~~&  $-0.2537$         ~~~~~~~~~~~~~~~~~~&$-0.2318$       \\
$r_c$    (fm )         ~~~~~~~~~~~~~~~~~~~&   0.337         ~~~~~~~~~~~~~~~~~~& 0.292        \\
\hline\hline
\end{tabular}
\end{center}
\end{table}

\begin{table*}[htb]
\begin{center}
\caption{The predicted bottom meson masses (MeV) compared with the data and some other model predictions. The mixing angle of $1^{3}P_{1}-1^{1}P_{1}$ and $1^{3}D_{2}-1^{1}D_{2}$ obtained in present work are $\theta_{1P}=-35.2^{\circ}$, and $\theta_{1D}=-39.5^{\circ}$. In the table, $\beta_{eff}^{NR}$ and $\beta_{eff}^R$ stand for the effective harmonic oscillator parameters (GeV) of our nonrelativistic quark model calculations and those with the relativized quark model calculations~\cite{Godfrey:2016nwn}, respectively, while $\beta_{eff}^{C}$ stands for our results including relativistic corrections of the length contraction effects.  }\label{massB}
\begin{tabular}{ccccccccccccccccc}\hline\hline
& State~~&$J^{P}$ &$\beta_{eff}^{NR}$~~&$\beta_{eff}^{C}$~~&$\beta_{eff}^R$~\cite{Godfrey:2016nwn} ~&Ours~  &KDR~\cite{Kher:2017mky} ~ &EFG~\cite{Ebert:2009ua} ~&LPW~\cite{Lu:2016bbk} &LL~\cite{Liu:2016efm}~&GI~\cite{Godfrey:2016nwn}~~
&AMS~\cite{Asghar:2018tha}~& PE~\cite{DiPierro:2001dwf}~&Exp.~\cite{Zyla:2020zbs}~\\
\hline %	
&$B(1 ^1S_{0})$~~&$0^{-}$~~&0.444 ~~&0.628 ~~&0.580 ~~&5279(fitted) ~~&5287    ~~&5280    ~~&5280    ~~&5273     ~~&5312    ~~&5268    ~~&5279    ~~&$5280$\\
&$B(1 ^3S_{1})$~~&$1^{-}$~~&0.418 ~~&0.575 ~~&0.542 ~~&5325(fitted) ~~&5323    ~~&5326    ~~&5329    ~~&5329     ~~&5371    ~~&5329    ~~&5324    ~~&$5325$\\
&$B(1 ^3P_{0})$~~&$0^{+}$~~&0.367 ~~&0.525 ~~&0.536 ~~&5722 ~~&5730    ~~&5749    ~~&5683    ~~&5776     ~~&5756    ~~&5704    ~~&5706    ~~&$\cdots$\\
&$B(1P_1)$     ~~&$1^{+}$~~&0.348 ~~&0.487 ~~&0.511 ~~&5716 ~~&5733    ~~&5723    ~~&5729    ~~&5719     ~~&5784    ~~&5739    ~~&5700    ~~&$5698$\\
&$B(1P'_1)$    ~~&$1^{+}$~~&0.348 ~~&0.487 ~~&0.499 ~~&5753(fitted) ~~&5752    ~~&5744    ~~&5754    ~~&5837     ~~&5777    ~~&5755    ~~&5742    ~~&$5726$\\
&$B(1 ^3P_{2})$~~&$2^{+}$~~&0.334 ~~&0.460 ~~&0.472 ~~&5727(fitted) ~~&5740    ~~&5741    ~~&5768    ~~&5739     ~~&5797    ~~&5769    ~~&5714    ~~&$5740$\\
&$B(2 ^1S_{0})$~~&$0^{-}$~~&0.327 ~~&0.472 ~~&0.477 ~~&5876 ~~&5926    ~~&5890    ~~&5910    ~~&5957     ~~&5904    ~~&5877    ~~&5886    ~~&$5863$\\
&$B(2 ^3S_{1})$~~&$1^{-}$~~&0.320 ~~&0.458 ~~&0.468 ~~&5899 ~~&5947    ~~&5906    ~~&5939    ~~&5997     ~~&5933    ~~&5905    ~~&5920    ~~&$\cdots$\\
&$B(1 ^3D_{1})$~~&$1^{-}$~~&0.316 ~~&0.453 ~~&0.488 ~~&6056 ~~&6016    ~~&6119    ~~&6095    ~~&6143     ~~&6110    ~~&6022    ~~&6025    ~~&$\cdots$\\
&$B(1D_2)$     ~~&$2^{-}$~~&0.312 ~~&0.445 ~~&0.463 ~~&5973 ~~&6031    ~~&6103    ~~&6004    ~~&5993     ~~&6095    ~~&6026    ~~&5985    ~~&$\cdots$\\
&$B(1D'_2)$    ~~&$2^{-}$~~&0.312 ~~&0.445 ~~&0.469 ~~&6067 ~~&6065    ~~&6121    ~~&6113    ~~&6165     ~~&6124    ~~&6031    ~~&6037    ~~&$\cdots$\\
&$B(1 ^3D_{3})$~~&$3^{-}$~~&0.312 ~~&0.445 ~~&0.444 ~~&5979 ~~&6016    ~~&6091    ~~&6014    ~~&6004     ~~&6106    ~~&6031    ~~&5993    ~~&$5971$\\
\hline\hline
\end{tabular}
\end{center}
\end{table*}

\begin{table*}[htb]
\begin{center}
\caption{The predicted bottom-strange meson masses (MeV) compared with the data and some other model predictions. The mixing angle of $1^{3}P_{1}-1^{1}P_{1}$ and $1^{3}D_{2}-1^{1}D_{2}$ obtained in present work are $\theta_{1P}=-39.6^{\circ}$, and $\theta_{1D}=-39.9^{\circ}$.
In the table, $\beta_{eff}^{NR}$ and $\beta_{eff}^R$ stand for the effective harmonic oscillator
parameters (GeV) obtained from our nonrelativistic quark model calculations and those with the relativized quark model calculations~\cite{Godfrey:2016nwn}, respectively, while $\beta_{eff}^{C}$ stands for our results including relativistic corrections of the length contraction effects.  }\label{massBs}
\begin{tabular}{ccccccccccccccccc}\hline\hline
& State~~&$J^{P}$ &$\beta_{eff}^{NR}$~~&$\beta_{eff}^{C}$~~&$\beta_{eff}^R$~\cite{Godfrey:2016nwn}~~&Ours~~ & KDR~\cite{Kher:2017mky}~~& EFG~\cite{Ebert:2009ua} &ZVR~\cite{Zeng:1994vj}~~&LPW~\cite{Lu:2016bbk}~~
&AMS~\cite{Asghar:2018tha}~~& GI~\cite{Godfrey:2016nwn}~~& PE~\cite{DiPierro:2001dwf}~~&Exp.\\
\hline %
&$B_{s}(1 ^1S_{0})$~~&$0^{-}$~~&0.507 ~~&0.681 ~~&0.636 ~~& 5367(fitted) &5367    ~~&5372    ~~ &5370     ~~&5362    ~~&5377    ~~&5394    ~~&5373    &$5367$ ~\cite{Zyla:2020zbs}  \\
&$B_{s}(1 ^3S_{1})$~~&$1^{-}$~~&0.475 ~~&0.621 ~~&0.595 ~~& 5415(fitted) &5413    ~~&5414    ~~ &5430     ~~&5413    ~~&5422    ~~&5450    ~~&5421    &$5416$ ~\cite{Zyla:2020zbs}\\
&$B_{s}(2 ^1S_{0})$~~&$0^{-}$~~&0.363 ~~&0.490 ~~&0.508 ~~& 5944 &6003    ~~&5976    ~~ &5930     ~~&5977    ~~&5929    ~~&5984    ~~&5985    &$\cdots$\\
&$B_{s}(2 ^3S_{1})$~~&$1^{-}$~~&0.356 ~~&0.465 ~~&0.497 ~~& 5966 &6029    ~~&5992    ~~ &5970     ~~&6003    ~~&5949    ~~&6012    ~~&6019    &$\cdots$\\
&$B_{s}(1 ^3P_{0})$~~&$0^{+}$~~&0.408 ~~&0.549 ~~&0.563 ~~& 5788 &5812    ~~&5833    ~~ &5750     ~~&5756    ~~&5770    ~~&5831    ~~&5804    &$\cdots$\\
&$B_{s}(1P_1)$     ~~&$1^{+}$~~&0.387 ~~&0.509 ~~&0.538 ~~& 5810 &5828    ~~&5831    ~~ &5790     ~~&5801    ~~&5801    ~~&5857    ~~&5805    &$\cdots$\\
&$B_{s}(1P'_1)$    ~~&$1^{+}$~~&0.387 ~~&0.509 ~~&0.528 ~~& 5821(fitted) &5842    ~~&5865    ~~ &5800     ~~&5836    ~~&5803    ~~&5861    ~~&5842    &$5829$~\cite{Zyla:2020zbs}\\
&$B_{s}(1 ^3P_{2})$~~&$2^{+}$~~&0.369 ~~&0.477 ~~&0.504 ~~& 5821(fitted) &5840    ~~&5842    ~~ &5820     ~~&5851    ~~&5822    ~~&5876    ~~&5820    &$5840$~\cite{Zyla:2020zbs}\\
&$B_{s}(1 ^3D_{1})$~~&$1^{-}$~~&0.351 ~~&0.472 ~~&0.504 ~~& 6101 &6119    ~~&6209    ~~ &6070     ~~&6142    ~~&6057    ~~&6182    ~~&6127    &$6114$~\cite{Aaij:2020hcw}\\
&$B_{s}(1D_2)$     ~~&$2^{-}$~~&0.344 ~~&0.459 ~~&0.487 ~~& 6061 &6128    ~~&6189    ~~ &6070     ~~&6087    ~~&6059    ~~&6169    ~~&6095    &$\cdots$~~~~\\
&$B_{s}(1D'_2)$    ~~&$2^{-}$~~&0.344 ~~&0.459 ~~&0.482 ~~& 6113 &6157    ~~&6218    ~~ &6080     ~~&6159    ~~&6064    ~~&6196    ~~&6140    &$6109$~\cite{Aaij:2020hcw}\\
&$B_{s}(1 ^3D_{3})$~~&$3^{-}$~~&0.340 ~~&0.451 ~~&0.467 ~~& 6067 &6172    ~~&6191    ~~ &6080     ~~&6096    ~~&6063    ~~&6179    ~~&6103    &$6064$~\cite{Aaij:2020hcw}\\

\hline\hline
\end{tabular}
\end{center}
\end{table*}

\section{mass spectrum}\label{spectrum}

To describe the bottom and bottom-strange meson spectra, we adopt a nonrelativistic linear
potential model. In this model, the effective potential is adopted as~\cite{Godfrey:1986wj,Barnes:2005pb,Eichten:1978tg,Godfrey:2004ya}
\begin{eqnarray}\label{H1}
V(r)=V_0(r)+V_{sd}(r),
\end{eqnarray}
where
\begin{eqnarray}\label{H0}
V_0(r)=-\frac{4}{3}\frac{\alpha_s}{r}+br+C_{0}
\end{eqnarray}
includes the standard color Coulomb interaction and linear confinement, and zero point energy $C_0$.
The spin-dependent part $V_{sd}(r)$ can be expressed as~\cite{Godfrey:1986wj,Godfrey:2004ya,Eichten:1980mw}
\begin{eqnarray}\label{H0}
V_{sd}(r)=H_{SS}+H_{T}+H_{LS},
\end{eqnarray}
where
\begin{eqnarray}\label{ss}
H_{SS}= \frac{32\pi\alpha_s}{9m_qm_{\bar{q}}}\tilde{\delta}_\sigma(r)\mathbf{S}_{q}\cdot \mathbf{S}_{\bar{q}}
\end{eqnarray}
is the spin-spin contact hyperfine potential. Here, we take $\tilde{\delta}_\sigma(r)=(\sigma/\sqrt{\pi})^3
e^{-\sigma^2r^2}$ as suggested in Ref.~\cite{Barnes:2005pb}. The tensor potential $H_T$ is adopted as
\begin{eqnarray}\label{t}
H_{T}= \frac{4}{3}\frac{\alpha_s}{m_qm_{\bar{q}}}\frac{1}{r^3}S_T,
\end{eqnarray}
with $S_T=\frac{3\mathbf{S}_{q}\cdot \mathbf{r}\mathbf{S}_{\bar{q}}\cdot \mathbf{r}}{r^2}-\mathbf{S}_{q}\cdot\mathbf{S}_{\bar{q}}$.

The spin-orbit interaction $H_{LS}$ can be decomposed
into symmetric part $H_{sym}$ and antisymmetric part
$H_{anti}$:
\begin{eqnarray}\label{vs}
H_{LS}=H_{sym}+H_{anti},
\end{eqnarray}
with
\begin{eqnarray}\label{vs}
H_{sym}=\frac{\mathbf{S_{+}\cdot L}}{2}\left[\left(\frac{1}{2m_{\bar{q}}^{2}}+\frac{1}{2m_{q}^{2}}\right)\left(\frac{4}{3}\frac{\alpha_{s}}{r^{3}}-\frac{b}{r}\right)
+\frac{8\alpha_{s}}{3m_{q}m_{\bar{q}}r^{3}}\right],\\
H_{anti}=\frac{\mathbf{S_{-}\cdot L}}{2}\left(\frac{1}{2m_{q}^{2}}-\frac{1}{2m_{\bar{q}}^{2}}\right)\left(\frac{4}{3}\frac{\alpha_{s}}{r^{3}}-\frac{b}{r}\right).\ \ \ \ \ \ \ \ \ \ \ \ \ \ \ \ \ \ \ \ \ \ \
\end{eqnarray}
In these equations, $\mathbf{L}$ is the relative orbital angular momentum of the $q\bar{q}$
system; $\mathbf{S}_q$ and $\mathbf{S}_{\bar{q}}$ are the spins of the quark $q$ and antiquark $\bar{q}$, respectively, and $\mathbf{S}_{\pm}\equiv\mathbf{S}_q\pm \mathbf{S}_{\bar{q}}$; $m_q$ and $m_{\bar{q}}$ are the
masses of quark $q$ and antiquark $\bar{q}$, respectively; $\alpha_s$ is
the running coupling constant of QCD; and $r$ is the distance between the quark $q$ and antiquark $\bar{q}$.
The six parameters in the above potentials ($\alpha_s$, $b$, $\sigma$, $m_q$, $m_{\bar{q}}$, $C_0$)
are determined by fitting the mass spectrum.

It should be emphasized that when $m_q\neq m_{\bar{q}}$, the antisymmetric
part of the spin-orbit potential, $H_{anti}$, can cause a configuration mixing between
spin triplet $n^{3}L_{J}$ and spin singlet $n^{1}L_{J}$. Thus, the physical states $nL_J$ and $nL'_J$ are expressed as
\begin{equation}\label{mixst}
\left(
  \begin{array}{c}
   nL_J\\
   nL'_J\\
  \end{array}\right)=
  \left(
  \begin{array}{cc}
   \cos\theta_{nL} &\sin\theta_{nL}\\
  -\sin\theta_{nL} &\cos\theta_{nL}\\
  \end{array}
\right)
\left(
  \begin{array}{c}
  n^{1}L_{J}\\
  n^{3}L_{J}\\
  \end{array}\right).
\end{equation}
where $J=L=1,2,3\cdots$, and the $\theta_{nL}$ is the mixing angle.
In this work $nL'_J$ corresponds to the higher mass mixed state
as often adopted in the literature.

In this work, we solve the radial Schr\"{o}dinger equation by using
the three-point difference central method~\cite{Haicai} from central ($r=0$)
towards outside ($r\to \infty$) point by point. This method was successfully to
deal with the spectroscopies of $c\bar{c}$, $b\bar{b}$, $b\bar{c}$ and $s\bar{s}$
~\cite{Deng:2016ktl,Deng:2016stx,Li:2019tbn,Li:2019qsg,Li:2020xzs}.
To overcome the singular behavior of $1/r^3$ in the spin-dependent potentials, following
the method of our previous works~\cite{Deng:2016ktl,Deng:2016stx,Li:2019tbn,Li:2019qsg,Li:2020xzs}, we introduce a cutoff distance
$r_c$ in the calculation. Within a small range $r\in (0,r_c)$, we let $1/r^3=1/r_c^3$.
By introducing the cutoff distance $r_c$, we can nonperturbatively include the corrections from these
spin-dependent potentials containing $1/r^3$ to both
the mass and wave function of a meson state, which are crucial
for our predicting the decay properties.

The model parameters adopted in this work are listed in Table~\ref{parameter}.
To be consistent with our previous study~\cite{Li:2019tbn,Li:2019qsg,Li:2020xzs,Liu:2019wdr}, the bottom quark mass $m_b$, the light up
or down quark mass $m_{u/d}$, the strange quark mass $m_s$ are taken
from the determinations, i.e., $m_b=4.852$ GeV, $m_{u/d}=0.45$ GeV, $m_{s}=0.60$ GeV.
The other four parameters ($\alpha_s$, $b$, $\sigma$, $C_0$) for the bottom
meson sector, they are determined by fitting the masses of the well established
states $B(5279)$, $B^*(5325)$, $B_1(5721)$ and $B^*_2(5747)$, while for the bottom-strange
meson sector, they are determined by fitting the masses of the well established
states $B_s(5367)$, $B^*_s(5415)$, $B_{s1}(5830)$ and $B^*_{s2}(5840)$. It should
be pointed out that the zero-point-energy parameter $C_0$ is taken to be zero
for the $c\bar{c}$, $b\bar{b}$, $b\bar{c}$ heavy quarkonium systems in the literature
~\cite{Deng:2016ktl,Deng:2016stx,Li:2019tbn,Li:2019qsg}.
For these heavy quarkonium systems, the zero point energy can be absorbed into
the constituent quark masses because it only affects the heavy quark masses slightly.
However, if the zero point energy is absorbed into the meson systems containing
light quarks, it can significantly change the light constituent quark masses,
which play an important role in the spin-dependent potentials.
Thus, to obtain a good description of both the masses and the hyperfine/fine splittings
for the meson systems containing light quarks, a zero-point-energy parameter
$C_0$ is usually adopted in the calculations.

\begin{figure}[htbp]
\begin{center}
\centering  \epsfxsize=8.5cm \epsfbox{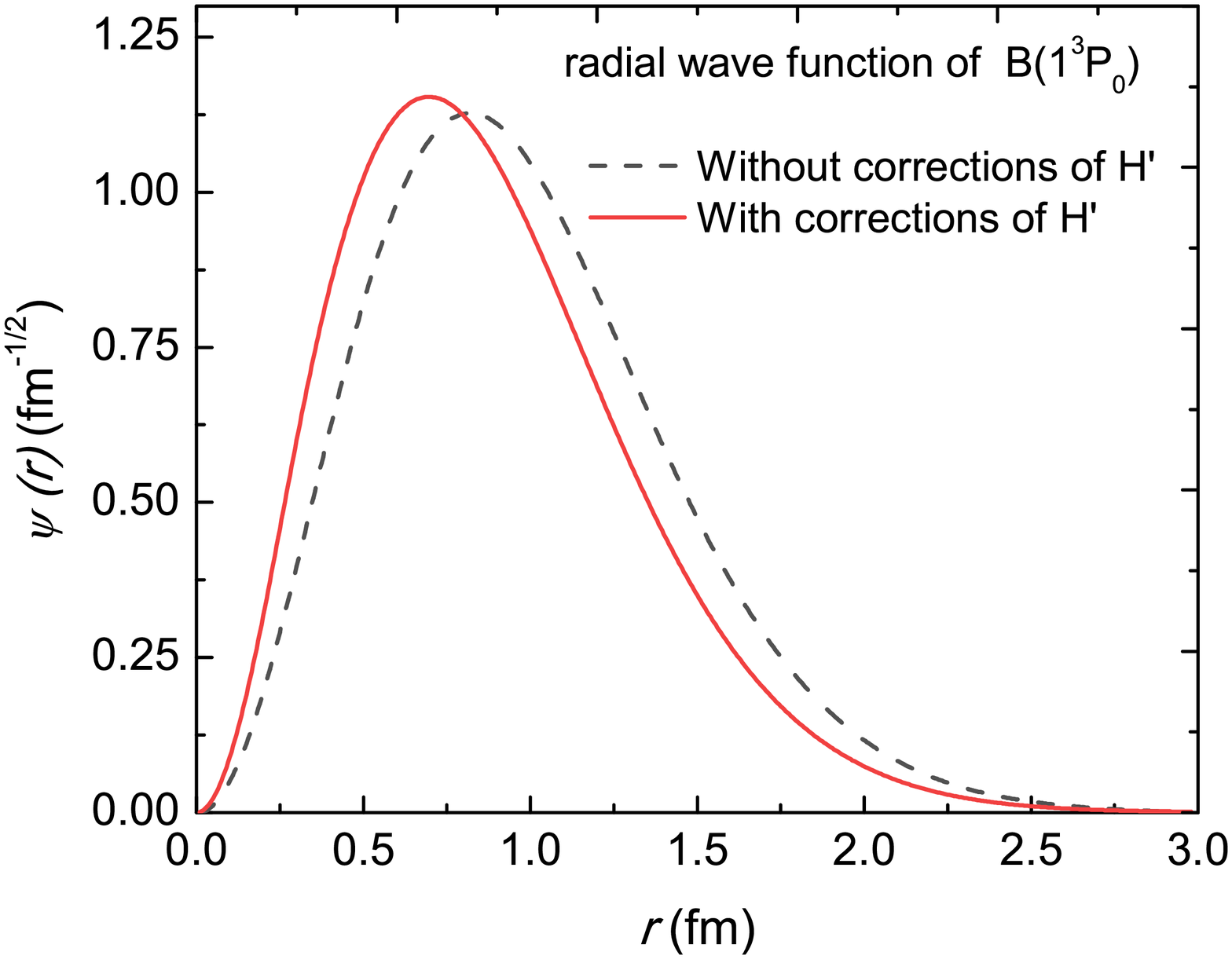}
\vspace{-0.4cm}\caption{The radial wave function $\psi(r)$ of the $B(1^3P_0)$ state. The solid and dashed lines stand
for the wave functions with and without corrections from the perturbation term $H'$ containing $1/r^{3}$, respectively.} \label{wf}
\end{center}
\end{figure}

\begin{figure*}[htbp]
\begin{center}
\centering  \epsfxsize=16cm \epsfbox{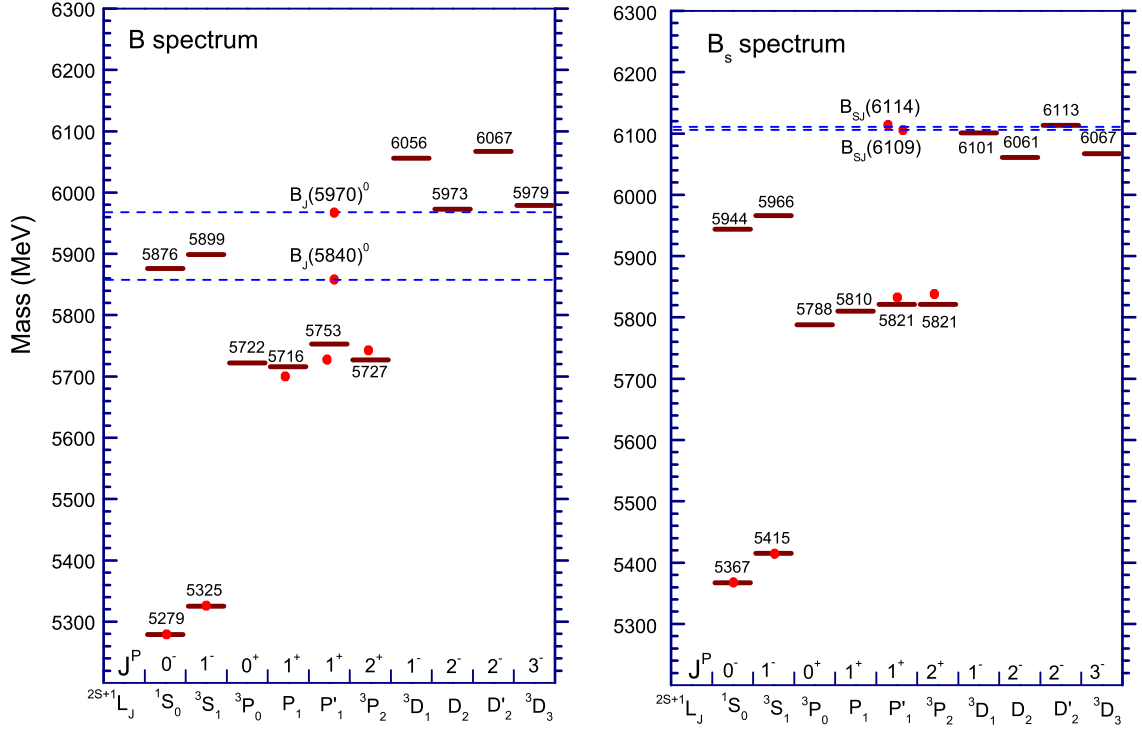}
\vspace{-0.8cm}\caption{The predicted mass spectra of $B$ and $B_s$ mesons. The solid circles stand for
the measured masses obtained from the Particle
Data Group~\cite{Zyla:2020zbs} and the recent LHCb measurements~\cite{Aaij:2020hcw}.} \label{mass}
\end{center}
\end{figure*}

Finally, we need determine the cutoff distance $r_c$ in our calculations.
It is found that the masses of the $1^3P_0$ states are more sensitive 
to the cutoff distance $r_c$. The reason is that the singular terms of $1/r^3$ in the 
spin-dependent potentials have more effects on the 
mass of the $1^3P_0$ state due to its relatively larger factors of $\langle\mathbf{S_{+}\cdot L}\rangle$
and $\langle S_T \rangle$ than the other excited meson states. 
Thus, in the present work, we determine the $r_c$ by fitting the masses of
the $B(1^3P_0)$ and $B_s(1^3P_0)$ states. Note that when the other parameters
are well determined, the masses of these $1^3P_0$ states can be reliably worked out
with the perturbation method without introducing the cutoff distance $r_c$, although the
wave functions obtain no corrections from the spin-dependent potentials containing $1/r^3$.
To calculate the masses of the $B(1^3P_0)$ and $B_s(1^3P_0)$ states with the perturbation method, we let $H=H_{0}+H'$,
where $H_0$ is the main part of the Hamiltonian without singular $1/r^{3}$ term contributions, i.e.,
\begin{eqnarray}
H_0&=&\frac{P^2}{2\mu}+V_0(r)
+\frac{32\pi\alpha_s}{9m_qm_{\bar{q}}}\tilde{\delta}_\sigma(r)\langle\mathbf{S}_{q}\cdot \mathbf{S}_{\bar{q}}\rangle \nonumber
\\&+&\frac{\langle \mathbf{S_{+}\cdot L}\rangle}{2}\left[\left(\frac{1}{2m_{\bar{q}}^{2}}+\frac{1}{2m_{q}^{2}}\right)\left(-\frac{b}{r}\right)\right],
\end{eqnarray}
where $\frac{P^2}{2\mu}$ is the kinetic energy term with a reduced mass $\mu=m_qm_{\bar{q}}/(m_q+m_{\bar{q}})$;
while $H'$ is the perturbation part containing $1/r^{3}$ terms, i.e.,
\begin{eqnarray}
H'&=&\frac{\langle\mathbf{S_{+}\cdot L}\rangle}{2}\left[\left(\frac{1}{2m_{\bar{q}}^{2}}+\frac{1}{2m_{q}^{2}}\right)\left(\frac{4}{3}\frac{\alpha_{s}}{r^{3}}\right)\nonumber
+\frac{8\alpha_{s}}{3m_{q}m_{\bar{q}}r^{3}}\right]
\\&+&\frac{4}{3}\frac{\alpha_s}{m_qm_{\bar{q}}}\frac{1}{r^3}\left\langle S_T\right\rangle.
\end{eqnarray}
For a $1^3P_0$ state, one can easily obtain the matrix elements $\langle\mathbf{S}_{q}\cdot \mathbf{S}_{\bar{q}}\rangle=1/4$,
$\langle\mathbf{S_{+}\cdot L}\rangle=-2$, and $\langle S_T \rangle=-1$ by using the relations
$\langle \mathbf{S}_q\cdot \mathbf{S}_{\bar{q}}\rangle= S(S+1)/2-3/4$,
$\langle \mathbf{S_+}\cdot \mathbf{L}\rangle = [J(J+1)-L(L+1)-S(S+1)]/2$, and
$\langle S_T \rangle =-[6\langle\mathbf{S_+}\cdot \mathbf{L}\rangle^2+3\langle\mathbf{S_+}\cdot \mathbf{L}\rangle
-2\mathbf{L}^2\mathbf{S_+}^2]/[2(2L-1)(2L+3)]$. By solving the radial Schr\"{o}dinger equation of
$H_{0} \psi^{(0)}(r) =E_{0} \psi^{(0)}(r) $ with the three-point difference central method,
we work out the eigen energies $E_{0}=508$, 415 MeV for the $B(1^3P_0)$ and $B_s(1^3P_0)$ states
as well as their zero order radial wave functions $\psi^{(0)}(r)$. Using these
zero order wave functions, we further work out the perturbation energies
$\langle\psi^{(0)}(r)|H'|\psi^{(0)}(r)\rangle=-88,-79$ MeV for the $B(1^3P_0)$ and $B_s(1^3P_0)$ states, respectively.
Finally, with the relation
$M=m_q+m_{\bar{q}}+E_{0}+\langle\psi^{(0)}(r)|H'|\psi^{(0)}(r)\rangle$ we predict
the masses 5722 and 5788 MeV for $B(1^3P_0)$ and $B_s(1^3P_0)$, respectively.
These masses calculated with the perturbation method are in good agreement with the predictions
in the literature~\cite{Kher:2017mky, Ebert:2009ua,Asghar:2018tha,DiPierro:2001dwf,Zeng:1994vj}.
Within a small range $r\in (0,r_c)$, letting $1/r^3=1/r_c^3$ in the perturbation
Hamiltonian $H'$, we solve the radial Schr\"{o}dinger equation
$(H_{0}+H') \psi(r) =E  \psi (r)$ with the three-point difference central method again. 
By reproducing the masses 5722 and 5788 MeV of
the $B(1^3P_0)$ and $B_s(1^3P_0)$ states obtained with the perturbation method, we
determine the cutoff distance parameters $r_c=0.337$ and 0.292 fm
for the $B$ and $B_s$ mass spectra, respectively. The corrections of the perturbation
part $H'$ to the radial wave function $\psi(r)$ can be conveniently included by
introducing the cutoff distance $r_c$. As an example, in Fig.~\ref{wf} we plot the radial wave
functions of the $B(1^3P_0)$ state with and without the corrections from
the perturbation part $H'$ containing $1/r^{3}$. It is seen that the
perturbation term $H'$ has an obvious correction to the radial wave
function.

With the determined model parameters listed in Table~\ref{parameter}, by solving the radial Schr\"{o}dinger equation
we obtain the masses of the bottom and bottom-strange meson states, which have been listed in Tab.~\ref{massB} and
Tab.~\ref{massBs}, respectively. For comparison, some other model predictions in Refs.~\cite{Kher:2017mky,Ebert:2009ua,Lu:2016bbk,Liu:2016efm,Godfrey:2016nwn,
Asghar:2018tha,DiPierro:2001dwf,Zeng:1994vj} and the data from the Review of Particle Physics (RPP) of Particle
Data Group (PDG)~\cite{Zyla:2020zbs} are listed in the same table as well.
Furthermore, for a clarity, the spectra are also shown in Fig.~\ref{mass}.
It is shown that the $B_J(5970)$ resonance and the structures $B_{sJ}(6064)$
and $B_{sJ}(6114)$ newly observed at LHCb can be explained as the $1D$-wave states from the point of view of the mass,  while
the $B_J(5840)$ may be a good candidate of the $2S$-wave state.

%Our predicted $B$-meson mass spectrum is comparable with
%that predicted in Refs.~\cite{DiPierro:2001dwf,Asghar:2018tha},
%while our predicted $B_s$-meson mass spectrum is comparable with
%that predicted in Refs.~\cite{Zeng:1994vj,Lu:2016bbk,Asghar:2018tha}.

\section{strong and radiative decays}\label{decay}

\subsection{Models}

We calculate the strong decays of the bottom and bottom-strange mesons with the
chiral quark model~\cite{Manohar:1983md,Li:1994cy,Li:1997gd,Zhao:2002id}.
In this model, the light pseudoscalar mesons, i.e. $\pi$, $K$ and $\eta$, are treated
as fundamental states. The low energy quark-pseudoscalar-meson and quark-vector-meson
interactions in the SU(3) flavor basis are described by the
effective Lagrangian
\begin{equation}\label{Hm}
H_m=\sum_{j}\frac{1}{f_m}\bar{\psi}_j\gamma_{\mu}\gamma_{5}\psi_j\vec{\tau}\cdot\partial^{\mu}\vec{\phi}_m,
\end{equation}
where $\psi_j$ represents the $j$th quark field in the
hadron, $\phi_m$ is the pseudoscalar meson field, $f_m$ is the
pseudoscalar meson decay constant. The nonrelativistic form of Eq.~(\ref{Hm}) is given by~\cite{Li:1994cy,Li:1997gd,Zhao:2002id}
\begin{eqnarray}\label{ccpk}
H_{m}=\sum_j\left[A \vsig_j \cdot \textbf{q}
+h\vsig_j\cdot \textbf{p}_j\right]I_j
\varphi_m,
\end{eqnarray}
in the center-of-mass system of the initial meson,  where we have
defined $A\equiv -(1+\frac{\omega_m}{E_f+M_f})$ and $h\equiv\frac{\omega_m}{2\mu_q}$. In Eq.~(\ref{ccpk}),  $\textbf{q}$ and $\omega_m$ are the
three-vector momentum and energy of the final-state light meson,
respectively; $\textbf{p}_j$ is the internal momentum operator of
the $j$th quark in the heavy-light meson rest frame; $\vsig_j$ is
the spin operator corresponding to the $j$th quark of the
heavy-light system; and $\mu_q$ is a reduced mass given by
$1/\mu_q=1/m_j+1/m'_j$ with $m_j$ and $m'_j$ for the masses of the
$j$th quark in the initial and final mesons, respectively. The plane wave part of the
emitted light meson is $\varphi_m=e^{-i\textbf{q}\cdot
\textbf{r}_j}$, and $I_j$ is the flavor operator defined for the
transitions in the SU(3) flavor space. The chiral quark model has been successfully
applied to describe the strong decays of the heavy-light mesons and baryons~\cite{DiPierro:2001dwf,Xiao:2014ura,Zhong:2010vq,Zhong:2008kd,
Zhong:2009sk,Liu:2012sj,Zhong:2007gp,Xiao:2013xi,Nagahiro:2016nsx}.
The details of this model can be found in Refs.~\cite{Zhong:2008kd,Zhong:2007gp}.
It should be mentioned that the chiral quark model is similar to the pseudoscalar
emission model in the literature~\cite{Godfrey:1985xj,Koniuk:1979vy,Capstick:2000qj}.
The nonrelativistic form of quark-pseudoscalar-meson interactions expressed in Eq.~(\ref{ccpk}) is similar to that of the pseudoscalar
emission model, except that the factors $A\equiv -(1+\frac{\omega_m}{E_f+M_f})$ and $h\equiv \frac{\omega_m}{2\mu_q}$ in this work have an explicit dependence on the energies of final hadrons.

Meanwhile, to treat the radiative decay of a hadron we apply
the constituent quark model~\cite{Brodsky:1968ea,Close:1970kt,Close:1989aj,Li:1994cy,Li:1993zzb,Li:1997gd,Zhao:1998fn,Zhao:2000tb,Zhao:2001jw,Zhao:2002id}.
In this model, the quark-photon EM coupling at the tree level
is adopted as
\begin{eqnarray}\label{he}
H_e=-\sum_j
e_{j}\bar{\psi}_j\gamma^{j}_{\mu}A^{\mu}(\mathbf{k},\mathbf{r}_j)\psi_j,
\end{eqnarray}
where $A^{\mu}$ represents the photon field with three-momentum $\mathbf{k}$. $e_j$ and $\mathbf{r}_j$
stand for the charge and coordinate of the constituent quark $\psi_j$, respectively.
In the initial-hadron-rest system, including the effects of the binding potential between
quarks, the nonrelativistic form of the quark-photon EM coupling can be written as~\cite{Brodsky:1968ea,Li:1994cy,Li:1997gd,Zhao:2002id,Zhao:1998fn,Zhao:2000tb,Zhao:2001jw,Li:1993zzb}
\begin{equation}\label{he2}
H_{e}^{nr}=\sum_{j}\left[e_{j}\mathbf{r}_{j}\cdot\veps-\frac{e_{j}}{2m_{j}
}\vsig_{j}\cdot(\veps\times\hat{\mathbf{k}})\right]\varphi_{\gamma} ,
\end{equation}
where $\vsig_j$ stand for the Pauli spin vector for the $j$th quark.
The vector $\veps$ is the polarization vector of the photon. The plane wave part of the
emitted light meson is $\varphi_{\gamma}=e^{-i\textbf{k}\cdot
\textbf{r}_j}$. The first and second terms in Eq.(\ref{he2}) are responsible for the electric and
magnetic transitions, respectively. The second term
in Eq.(\ref{he2}) is the same as that used in Refs.~\cite{Godfrey:1985xj,Koniuk:1979vy,Capstick:2000qj,Copley:1969ft,Sartor:1986sf}, while the first term in Eq.(\ref{he2}) is different from $(1/m_j) \mathbf{p}_j\cdot\veps$ of these works
due to the binding potential effects are included~\cite{Brodsky:1968ea}.
This model has been successfully applied to describe the radiative decays of baryon states~\cite{Yao:2018jmc,Wang:2017kfr,Lu:2017meb,Xiao:2017udy,
Wang:2017hej,Liu:2019vtx,Liu:2019wdr} and meson systems~\cite{Deng:2016stx,
Deng:2016ktl,Li:2019tbn,Close:2002ky,Chen:2020jku}.

For a strong decay process, the partial decay width is calculated with~\cite{Zhong:2008kd, Zhong:2007gp}
\begin{equation}\label{dww}
\Gamma_m=\left(\frac{\delta}{f_m}\right)^2\frac{(E_f+M_f)|\textbf{q}|}{4\pi
M_i(2J_i+1)} \sum_{J_{fz},J_{iz}}|\mathcal{M}_{J_{fz},J_{iz}}|^2 ,
\end{equation}
while for a radiative decay process, the partial decay width is calculated with~\cite{Deng:2016stx,Deng:2016ktl}
\begin{equation}\label{dww}
\Gamma_\gamma=\frac{|\mathbf{k}|^2}{\pi}\frac{2}{2J_i+1}\frac{M_{f}}{M_{i}}\sum_{J_{fz},J_{iz}}|\mathcal{A}_{J_{fz},J_{iz}}|^2,
\end{equation}
where $\mathcal{M}_{J_{fz},J_{iz}}$ and $\mathcal{A}_{J_{fz},J_{iz}}$ correspond to
the strong and radiative transition amplitudes, respectively.
The quantum numbers $J_{iz}$ and $J_{fz}$ stand for the third components of the total
angular momenta of the initial and final hadron states,
respectively. $\delta$ as a global parameter accounts for the
strength of the quark-meson couplings. It has been determined in our previous study of the strong
decays of the charmed baryons and heavy-light mesons
\cite{Zhong:2007gp,Zhong:2008kd}. Here, we fix its value the same as
that in Refs.~\cite{Zhong:2008kd,Zhong:2007gp}, i.e. $\delta=0.557$.

\subsection{Parameters}

In the calculation, the constituent quark masses for the $u$,
$d$, and $s$ quarks are taken with $m_u=m_d=450$ MeV and
$m_s=600$ MeV to be consistent with the spectrum study in Sec.~\ref{spectrum}.
The decay constants for $\pi$, $K$ and $\eta$ mesons are taken as $f_{\pi}=132$ MeV,
$f_K=f_{\eta}=160$ MeV, respectively. The masses of the well established hadrons involving in the
calculations are adopted from the PDG~\cite{Zyla:2020zbs}. The masses of the missing $B$- and $B_s$-meson
states are adopted our determinations by solving the Schr\"{o}inger equation in Sec.~\ref{spectrum}.

It should be mentioned that, we do not
directly adopt the numerical wave functions of $B$- and $B_s$-meson states
calculated by solving the Schr\"{o}inger equation. For simplicity,
we first fit them with a simple harmonic oscillator wave function
by reproducing the root-mean-square radius $\sqrt{\langle r^2 \rangle}$.
The obtained effective harmonic oscillator parameters $\beta_{eff}$
for the meson states are listed in Tab.~\ref{massB} and Tab.~\ref{massBs}.
It is found that the effective harmonic oscillator
parameters $\beta_{eff}^{NR}$ obtained from our nonrelativistic quark
model calculations are obviously smaller than the parameters $\beta_{eff}^{R}$ obtained from
the relativized quark models~\cite{Godfrey:2016nwn}. It indicates
that the relativistic effects of the length contraction on the wave function may be important.
To take into account the relativistic effects,
as that suggested in Ref.~\cite{Licht:1970pe} we introduce the Lorentz boost factor $\gamma$ in the spatial wave function, i.e.,
\begin{equation}\label{eqss}
\psi_{nlm}(\mathbf{r})\to \psi_{nlm}(\gamma \mathbf{r}),
\end{equation}
where $\gamma =M_q/E_q$. $M_q$ and $E_q$ correspond to the effective mass
and energy of the light quark, respectively.  According to Ref.~\cite{Jaczko:1998uj},
the effective mass $M_q$ can be estimated by $M_q=\sqrt{\langle p^2\rangle+m_q^2}$,
while the energy $E_q$ is estimated by $E_{q}=\langle p\rangle^2/(2M_q)+M_q$.
To realize this transformation, we only need replace $\beta_{eff}^{NR}$ in the harmonic oscillator
wave function with $\beta_{eff}^C=\gamma \beta_{eff}^{NR}$. The effective harmonic oscillator
parameters $\beta_{eff}^C$ including relativistic corrections of the length contraction are given in Tab.~\ref{massB} and
Tab.~\ref{massBs} as well. It is found that the $\beta_{eff}^C$ values are comparable with
those obtained with the relativized quark models~\cite{Godfrey:2016nwn}.
In Refs.\cite{Zhong:2008kd,Xiao:2014ura}, the strong decays of the heavy-light
meson states are studied with the chiral quark model by using the simple harmonic oscillator
wave functions with fixed harmonic oscillator parameters $\beta=468,466$ MeV for
the $B$ and $B_s$ spectra, respectively, which are close to the parameters $\beta_{eff}^C$ determined
for the $1P$-, $1D$- and $2S$-wave states in present work. The parameters $\beta_{eff}^C$ of
the ground states $1^1S_0$ and $1^3S_1$ are notably lager than
that of the excited states, this effect is mainly caused by
the strong color Coulomb interaction at the small distance $r$ between two quarks.

The effective parameters $\beta_{eff}^C$ of the ground states $B(1^1S_0)$ and $B^*(1^3S_1)$
are crucial for understanding the decay properties of the excited $B$ and $B_s$ states
because all of the excited states should decay into these ground states.
Considering the uncertainty of the parameters $\beta_{eff}^C$ of the ground
states $B(1^1S_0)$ and $B^*(1^3S_1)$, we properly adjust their $\beta_{eff}^C$ parameters
to more reasonably describe the strong decay properties of the well established $1P$-wave states $B_2^*(5747)^0$ and $B_{s2}^*(5840)$.
In this work, we determine them to be $\beta_{B(1^1S_0)}=0.537$ GeV and $\beta_ {B^*(1^3S_1)}=0.510$ GeV
for $B(1^1S_0)$ and $B^*(1^3S_1)$, respectively. There is about a $10\%$ correction
to the effective parameters $\beta_{eff}^C$.

With above parameters, our calculated decay properties for the $1P$-, $2S$-, and $1D$-wave states
are listed in Tables~\ref{decayp}, ~\ref{decayB2s}, and ~\ref{decay2d1}, respectively.
From Table~\ref{decayp}, it is found that the decay properties of the
well-established $1P$-wave state can be successfully described.

\section{discussion}\label{DIS}

\begin{table*}[htb]%£¨except 11S0£©
\caption{Partial and total decay widths (MeV) for $1P$-wave bottom and bottom-strange mesons compared with the the data and some recent model  predictions. It should be mention that some masses for the initial states adopted in the literature are slightly different.
The total widths inside the square brackets are estimated with the mixing angle $\theta_{1P}=-(55\pm 5)^\circ$.   }\label{decayp}
\begin{tabular}{cccccccccccccccccccccc}  \midrule[1.0pt]\midrule[1.0pt]
 $n^{2S+1}L_J$  ~~& State ~~& Channel ~~& ~~Ours
  ~~& XZ~\cite{Zhong:2008kd} ~~& SSC~\cite{Sun:2014wea} ~~&~~~~ LPW~\cite{{Lu:2016bbk}} ~~~~&~~~GI~\cite{Godfrey:2016nwn}~~~~&~~~AMS~\cite{Asghar:2018tha} & $\Gamma_{exp}$~\cite{Zyla:2020zbs}
 \\
\midrule[1.0pt]	                                                   % ME               zhong     liu           lv      GI      ANM           exp
$1^3P_0$     ~&$B^*_0(5722)^+$    ~&   $B^+\pi^0+B^0\pi^+$        ~&  97.5+195.5           ~&         ~&         ~&         ~&                       & 141.5 &                     \\
             ~&                   ~&   $B^{*+}\gamma$             ~&  477$\times10^{-3}$   ~&         ~&         ~&                      ~& 325$\times10^{-3}$   & 575$\times10^{-3}$  &                     \\
             ~&                   ~&   Total                      ~&  293                  ~&         ~&         ~&                      ~&       & 142.08  &                     \\
             ~&$B^*_0(5722)^0$    ~&   $B^0\pi^0+B^+\pi^-$        ~&  97.5+195.3~&  272   ~&  225    ~& 230.43  ~& 154                   & 141.5 &                     \\
             ~&                   ~&   $B^{*0}\gamma$             ~&  149$\times10^{-3}$   ~&         ~&         ~& 116.9$\times10^{-3}$ ~& 92.7$\times10^{-3}$  & 175$\times10^{-3}$  &                     \\
             ~&                   ~&   Total                      ~&  292.8                ~&  272    ~&  225    ~& 230.43               ~& 154   & 141.7 &                     \\
$1^3P_2$     ~&$B^*_2(5747)^+$    ~&   $B^+\pi^0+B^0\pi^+$        ~&  5.3+10.6             ~&         ~&         ~&                      ~&       &9.77  &                     \\
             ~&                   ~&   $B^{*+}\pi^0+B^{*0}\pi^+$  ~&  5.1+9.9              ~&         ~&         ~&                      ~&       &9.79  &                     \\
             ~&                   ~&   $B^{*+}\gamma$             ~&  146$\times10^{-3}$   ~&         ~&         ~&                      ~& 444$\times10^{-3}$   &761$\times10^{-3}$   &                     \\
             ~&                   ~&   Total                      ~&  31                   ~&         ~&         ~&                      ~& 11.71 &20.3  & $20\pm5$       \\
             ~&$B^*_2(5747)^0$    ~&   $B^0\pi^0+B^+\pi^-$        ~&  5.5+10.8             ~&  25     ~&  1.9    ~& 12.62                ~& 6.23  &9.77  &                     \\
             ~&                   ~&   $B^{*0}\pi^0+B^{*+}\pi^-$  ~&  5.2+10.2             ~&  22     ~&  1.8    ~& 11.89                ~& 5.04  &9.79  &                     \\
             ~&                   ~&   $B^{*0}\gamma$             ~&  51$\times10^{-3}$    ~&         ~&         ~&177.7$\times10^{-3}$  ~& 126$\times10^{-3}$   &232$\times10^{-3}$  &                     \\
             ~&                   ~&   Total                      ~&  \textbf{32} (fitted)                  ~&  47     ~&  3.7    ~& 24.51                ~& 11.40 &19.8 &$24.2\pm1.7$         \\
$1P   $      ~&$B_1(5680)^+$      ~&   $B^{*+}\pi^0+B^{*0}\pi^+$  ~&  46.8+93.9            ~&         ~&         ~&                      ~& 163   &125.53&                     \\
             ~&                   ~&   $B^{*+}\gamma$             ~&  75$\times10^{-3}$    ~&         ~&         ~&                      ~& 300   &448   &                     \\
             ~&                   ~&   $B^+  \gamma$              ~&  111$\times10^{-3}$   ~&         ~&         ~&                      ~& 132$\times10^{-3}$   &415$\times10^{-3}$  &                     \\
             ~&                   ~&   Total                      ~&  140.9                ~&         ~&         ~&                      ~& 163   &126.4  & $128\pm18$          \\
             ~&$B_1(5680)^0$      ~&   $B^{*0}\pi^0+B^{*+}\pi^-$  ~&  46.8+93.9  ~&  219   ~&  200    ~& 199.4   ~& 163   &125.53&                     \\
             ~&                   ~&   $B^{*0}\gamma$             ~&  24$\times10^{-3}$    ~&         ~&         ~& 53.1$\times10^{-3}$  ~& 85.5$\times10^{-3}$  &137$\times10^{-3}$   &                     \\
             ~&                   ~&   $B^0  \gamma$              ~&  38$\times10^{-3}$    ~&         ~&         ~& 130.2$\times10^{-3}$   ~& 37.8$\times10^{-3}$  &127$\times10^{-3}$   &                     \\
             ~&                   ~&   Total                      ~&  140.8                ~&  219    ~&  200    ~& 199.4   ~& 163   &125.8 &$128\pm18$           \\
$1P'   $     ~&$B_1(5721)^+$      ~&   $B^{*+}\pi^0+B^{*0}\pi^+$  ~&  13.8+27.4            ~&         ~&         ~&         ~& 6.80  & 15.62&                     \\
             ~&                   ~&   $B^{*+}\gamma$             ~&  206$\times10^{-3}$   ~&         ~&         ~&         ~& 300$\times10^{-3}$   & 339$\times10^{-3}$  &                     \\
             ~&                   ~&   $B^+  \gamma$              ~&  69$\times10^{-3}$    ~&         ~&         ~&         ~& 132$\times10^{-3}$   & 448$\times10^{-3}$  &                     \\
             ~&                   ~&   Total                      ~&  41.4 [$24.5\pm 2.5$] ~&         ~&         ~&         ~& 7.27  & 16.4 &  $31\pm6$         \\
             ~&$B_1(5721)^0$      ~&   $B^{*0}\pi^0+B^{*+}\pi^-$  ~&  13.8+27.4            ~&  30     ~&  10     ~& 40.63   ~& 6.80  & 15.62 &                     \\
             ~&                   ~&   $B^{*0}\gamma$             ~&  66$\times10^{-3}$    ~&         ~&         ~&108.5$\times10^{-3}$    ~& 27.8$\times10^{-3}$  & 103$\times10^{-3}$  &                     \\
             ~&                   ~&   $B^0  \gamma$              ~&  24$\times10^{-3}$    ~&         ~&         ~&60.4$\times10^{-3}$     ~& 106$\times10^{-3}$   & 137$\times10^{-3}$  &                     \\
             ~&                   ~&   Total                      ~&  41.3 [$24.5\pm 2.5$] ~&  30     ~&  10     ~& 40.63   ~& 6.93  & 15.9 &$27.5\pm3.4$         \\
$1^3P_0$     ~&$B_{s0}^*(5788)^0$ ~&   $B^+K^-+B^0\bar{K^0}$      ~&  86.7+76.7          ~&  227    ~&  225    ~&         ~& 138   &135.66  &                     \\
             ~&                   ~&   $B_s^{*0}\gamma$           ~&  102$\times10^{-3}$   ~&         ~&         ~&84.7$\times10^{-3}$  ~& 76$\times10^{-3}$    &133$\times10^{-3}$  &                     \\
             ~&                   ~&   Total                      ~&  163.4                  ~&  227    ~&  225    ~&         ~& 138   &135.8  &                     \\
$1^3P_2$     ~&$B_{s2}^*(5840)^0$ ~&   $B^+K^-+B^0\bar{K^0}$      ~&  0.69+0.61            ~&  2      ~&         ~&   1.9   ~&0.663  &1.55      &                     \\
             ~&                   ~&   $B^{*+}K^-+B^{*0}\bar{K^0}$~&  0.06+0.04               ~&  0.12   ~&         ~&   0.14  ~&0.00799&0.13   &                     \\
             ~&                   ~&   $B_s^{*0}\gamma$           ~&  51$\times10^{-3}$    ~&         ~&         ~&   159$\times10^{-3}$  ~&106$\times10^{-3}$    &225$\times10^{-3}$  &                     \\
             ~&                   ~&   Total                      ~&  \textbf{1.31} (fitted)                ~&  2      ~&  0.26   ~&   1.66  ~&0.777  &1.9      &1.49$\pm$0.27        \\
$1P   $      ~&$B_{s1}(5820)^0$   ~&   $B^{*+}K^-+B^{*0}\bar{K^0}$~&  $\cdots$             ~&  149    ~&  120    ~&         ~&       &      &                     \\
             ~&                   ~&   $B_s^{*0}\gamma$           ~&  56$\times10^{-3}$    ~&         ~&         ~& 39.5$\times10^{-3}$  ~&57.3$\times10^{-3}$   &      &                     \\
             ~&                   ~&   $B_s^0  \gamma$            ~&  37$\times10^{-3}$    ~&         ~&         ~&97.7$\times10^{-3}$  ~&47.8$\times10^{-3}$   &      &                     \\
             ~&                   ~&   Total                      ~&  $0.093$              ~&  149    ~&  120    ~&  160    ~&0.1051 &      &                     \\
$1P'   $     ~&$B_{s1}(5830)^0$   ~&   $B^{*+}K^-+B^{*0}\bar{K^0}$~&  4.3+3.3              ~&  $0.4-1$~&  $\sim0$~&         ~&       &      &                     \\
             ~&                   ~&   $B_s^{*0}\gamma$           ~&  53$\times10^{-3}$    ~&         ~&         ~& 98.8$\times10^{-3}$    ~& 36.9$\times10^{-3}$  &      &                     \\
             ~&                   ~&   $B_s^0  \gamma$            ~&  27$\times10^{-3}$    ~&         ~&         ~& 56.6$\times10^{-3}$    ~& 70.6$\times10^{-3}$  &      &                     \\
             ~&                   ~&   Total                      ~&  7.6 [$0.1-0.8$]      ~&  $0.4-1$~& $\sim0$ ~&  20     ~& 0.1075&      &0.5$\pm$0.3$\pm$0.3          \\

\midrule[1.0pt]\midrule[1.0pt]
\end{tabular}
\end{table*}

\subsection{$1P$-wave states}

Two $1P$-wave excited $B$ meson states $B_1(5721)$ and $B_2^*(5747)^{+,0}$
together with their flavor partners $B_{s1}(5830)$ and $B_{s2}^*(5840)^0$
in the $B_s$-meson family have been well-established in experiments~\cite{Zyla:2020zbs}. However,
two resonances with $J^P=0^+$, $B(1^3P_0)$ and $B_s(1^3P_0)$, and
two resonances with $J^P=1^+$, $B(P_1)$ and $B_s(P_1)$, predicted in
the quark model are still missing.

\subsubsection{$1^3P_2$ states}

There are no puzzles to assign the $B _2^*(5747)^{+,0}$ and $B_{s2}^*(5840)^0$
resonances to the $1^3P_2$ states in the $B$ and $B_s$ families, respectively.

In the $B$ meson sector, as the $1^3P_2$ state both the mass and width
of $B_2^*(5747)$ can be resonantly reproduced in the quark model.
Our fitted mass $M\simeq 5727$ MeV and width $\Gamma\simeq 31$ MeV
are compatible with the measurements $M_{exp}=5737$ MeV and width
$\Gamma_{exp}=(20\pm5)$ MeV for $B_2^*(5747)^+$.
This state dominantly decays into $B\pi$ and $B^*\pi$ channels with comparable partial widths.
The ideal partial width ratios between $B\pi$ and $B^*\pi$ channels for the $B_2^*(5747)^{+,0}$ are fitted to be
\begin{eqnarray}\label{5747}
R_1=\frac{\Gamma [B_2^*(5747)^{0}\to B^{*+}\pi^-]}{\Gamma [B_2^*(5747)^{0}\to B^+\pi^-]}\approx 0.94,\\
R_2=\frac{\Gamma [B_2^*(5747)^{+}\to B^{*0}\pi^+]}{\Gamma [B_2^*(5747)^{+}\to B^0\pi^+]}\approx 0.93,
\end{eqnarray}
which are also consistent with the recent LHCb measurements $R_1^{exp}=0.71\pm 0.14\pm 0.30$
and $R_2^{exp}=1.0\pm 0.5\pm 0.8$~\cite{Aaij:2015qla} and the
predictions in Refs.~\cite{Alhendi:2015rka,Sun:2014wea,Orsland:1998de,Ferretti:2015rsa,Gupta:2018xds,Asghar:2018tha,Yu:2019iwm,Lu:2016bbk}.

We further study the radiative decay processes
of $B_2^*(5747)^{+,0}\to B^{*+,0}\gamma$. Their partial decay widths
are predicted to be
\begin{eqnarray}\label{ra}
\Gamma [B_2^*(5747)^{0}\to B^{*0}\gamma]=51 \ \mathrm{keV},\\
\Gamma [B_2^*(5747)^{+}\to B^{*+}\gamma]=146 \ \mathrm{keV},
\end{eqnarray}
which are in good agreement with the predictions in Ref.~\cite{Orsland:1998de},
however, notably smaller than the predictions in Refs.~\cite{Asghar:2018tha,Godfrey:2019cmi,Yu:2019sqp,Godfrey:2016nwn,Lu:2016bbk}.
The radiative decay branching fractions can reach up to $\mathcal{O}(10^{-3})$,
thus radiative decays of $B_2^*(5747)^{+,0}\to B^{*+,0}\gamma$ might be observed
in future experiments.

In the $B_s$ meson sector, as the $1^3P_2$ state both the mass and width
of $B_{s2}^*(5840)$ can be well reproduced in the quark model as well.
Our fitted mass $M\simeq 5821$ MeV and width $\Gamma\simeq 1.3$ MeV
are compatible with the measured mass $M_{exp}=5840$ MeV and width
$\Gamma_{exp}=(1.49\pm 0.27)$ MeV.
There are two OZI allowed two-body
strong decay channels $BK$ and $B^*K$. The $BK$ mode governs the decays of $B_{s2}^*(5840)$.
The ideal partial width ratio between $B^*K$ and $BK$ is fitted to be
\begin{eqnarray}\label{5747}
R=\frac{\Gamma [B_{s2}^*(5840)\to B^{*+}K^-]}{\Gamma [B_{s2}^*(5840)\to B^{+}K^-]}\approx 8.7\%,
\end{eqnarray}
which is in good agreement with the recent LHCb measured one $R^{exp}=(9.3\pm 2.5)\%$~\cite{Aaij:2012uva}
and the predictions in the literature~\cite{Luo:2009wu,Sun:2014wea,Godfrey:2019cmi,Asghar:2018tha,Lu:2016bbk}.

Furthermore, it is found that $B_{s2}^*(5840)^0$ has a large decay rate into $B_s^*\gamma$, the
partial width and branching fraction are predicted to be
\begin{eqnarray}\label{5747}
\Gamma [B_{s2}^*(5840)^0\to B^{*0}_s\gamma] &\simeq & 51\ \mathrm{keV},\\
Br[B_{s2}^*(5840)^0\to B^{*0}_s\gamma]& \simeq & 3.4\%.
\end{eqnarray}
Our predicted radiative partial
decay width is consistent with those predictions in Refs.~\cite{Orsland:1998de,Yu:2019sqp}.
In the literature, a larger partial width $\Gamma[B_{s2}^*(5840)^0\to B^{*0}_s\gamma]\simeq 100-230$ keV
is predicted~\cite{Lu:2016bbk,Godfrey:2016nwn,Asghar:2018tha}.
The $B^{*0}_s\gamma$ decay channel of $B_{s2}^*(5840)^0$ may have good potentials to
be observed in future experiments.

\subsubsection{$1^3P_0$ states}

The $1^3P_0$ states in the $B$ and $B_s$ families are still missing experimentally.
In the $B$ meson sector, we predict that the mass of the $B(1^3P_0)$
state is $5722$ MeV, which is consistent with those predictions in Refs.~\cite{Gregory:2010gm,Kher:2017mky,Ebert:2009ua,Godfrey:2016nwn,DiPierro:2001dwf,Asghar:2018tha}.
The $B\pi$ channel is the only OZI-allowed
two body decay channel. Taking the estimated mass
$M=5722$ MeV, we obtain a broad width $\Gamma\simeq 290$ MeV
for the $B(1^3P_0)$ state, which is compatible with
our previous result $\Gamma\simeq 270$ MeV predicted with the SHO wave functions
in Ref.~\cite{Zhong:2008kd}. The $B(1^3P_0)$ state is also predicted to
be a broad state with a width of $\sim 150-250$ MeV in the other models
~\cite{Lu:2016bbk,Godfrey:2019cmi,Godfrey:2016nwn,Yu:2019iwm,Sun:2014wea,Zhu:1998wy}.
We also study the radiative decays of $B(1^3P_0)^{+,0}$,
our results have been listed in Table~\ref{decayp}, the predicted partial width for
$\Gamma[B(1^3P_0)^{+}\to B^{*+}\gamma]\simeq 477$ keV is about a factor 3
larger than that for $\Gamma[B(1^3P_0)^{0}\to B^{*0}\gamma]\simeq 149$ keV.
Our predictions are comparable with those predicted in Refs.~\cite{Asghar:2018tha,Godfrey:2016nwn}.
It should be mentioned that the radiative decays of $B(1^3P_0)^{+,0}$ are governed by
the $E1$ transitions. For a heavy-light meson system, the $E1$ transition
amplitude $\mathcal{A}$ is mainly contributed by
the light quark, it is proportional to $e_q/m_q$, where $e_q$ and $m_q$ stands for the charge and
mass of the light quark, respectively. The heavy quark contributions are strongly suppressed by
the heavy quark mass $m_Q$ (i.e., the factor $1/m_Q$). If neglecting the
heavy quark contributions, one has $\Gamma[B(1^3P_0)^{+}\to B^{*+}\gamma]:$
$\Gamma[B(1^3P_0)^{0}\to B^{*0}\gamma]=4:1$, which is slightly larger than the
ratio $\sim 3$ including heavy quark contributions.

In the $B_s$ meson sector, the mass of $B_s(1^3P_0)$ is
expected to be around $5788$ MeV, which is compatible with those predictions in Refs.~\cite{Gregory:2010gm,Godfrey:2016nwn,Kher:2017mky,Ebert:2009ua,Zeng:1994vj,Lu:2016bbk,Asghar:2018tha,DiPierro:2001dwf}.
The $BK$ channel is the only OZI-allowed
two body strong decay channel. Taking the estimated mass
$M=5788$ MeV, we obtain a broad width $\Gamma\simeq 270$ MeV
for the $B_s(1^3P_0)$ state. This prediction is compatible with
our previous result $\Gamma\simeq 227$ MeV predicted with the SHO wave functions
in Ref.~\cite{Zhong:2008kd}. The $B_s(1^3P_0)$ state is also predicted to
be a broad state with a width of $\sim 130-230$ MeV in the other models
~\cite{Godfrey:2019cmi,Godfrey:2016nwn,Sun:2014wea}.
We also study the radiative decay of $B_s(1^3P_0)$, the predicted partial width for
$\Gamma[B_s(1^3P_0)\to B_s^{0*}\gamma]\simeq 100$ keV is close to
the predictions in Refs.~\cite{Godfrey:2016nwn,Asghar:2018tha}. Finally,
it should be mentioned that in some works~\cite{Lu:2016bbk,Zeng:1994vj,Asghar:2018tha}, the predicted mass
of $B_s(1^3P_0)$ is below the $BK$ mass threshold, which will lead to a very narrow width for the $B_s(1^3P_0)$ state.

\begin{figure}[htbp]
\begin{center}
\centering  \epsfxsize=8.5cm \epsfbox{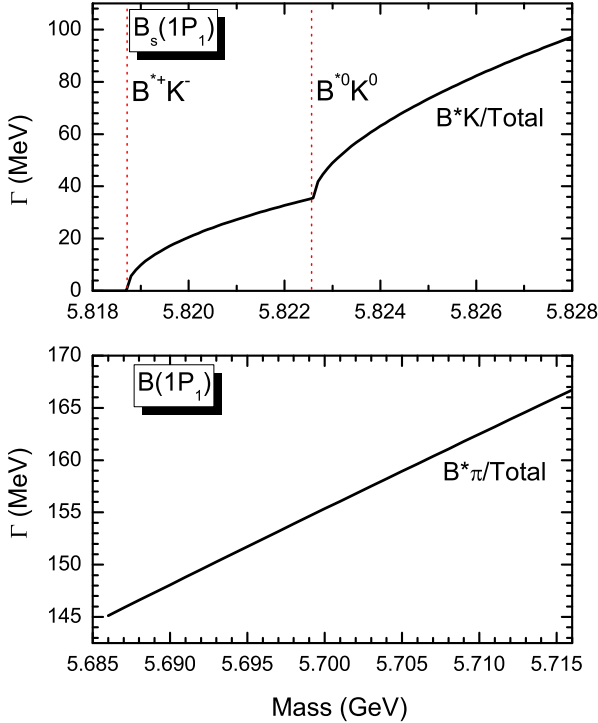}
\vspace{-0.4cm}\caption{Decay widths of the $1P$-wave mixed states $B_s(1P_1)$ and $B(1P_1)$
as a function of mass. The mixing angles for the $B_s(1P_1)$ and $B(1P_1)$
states are adopted the potential model predictions $\theta_{1P}=-39.6^{\circ}$ and $-35.2^{\circ}$, respectively.} \label{b1p}
\end{center}
\end{figure}

\subsubsection{ $1^1P_1$-$1^3P_1$ mixing}

The spin-orbit potential causes a strong configuration mixing between $1^{3}P_{1}$ and $1^{1}P_{1}$.
It is generally believed that $B_1(5721)$ and $B_{s1}(5830)$
correspond to the mixed states $|1P'_1\rangle$ via the $1^1P_1$-$1^3P_1$ mixing
in the $B$ and $B_s$ families, respectively. The other two mixed states
$B_{1}(1P_1)$ and $B_{s1}(1P_1)$ in the the $B$ and $B_s$ families are
waiting to be established in future experiments.

Considering $B_1(5721)$ as the mixed state $|1P'_1\rangle$ defined
in Eq.(\ref{mixst}), our fitted mass $M=5753$ MeV is reasonably comparable with the
measured value $M_{exp}=5726$ MeV. With the mixing angle
$\theta_{1P}=-35.2^{\circ}$ determined from our quark model,
the width of $B_1(5721)^0$ is predicted to be $\Gamma\simeq 41$ MeV, which is
in good agreement with the predictions in Refs.~\cite{Lu:2016bbk,Yu:2019iwm}, while slightly larger than the observed width
$\Gamma_{exp}\simeq 30$ MeV at LHC~\cite{Aaij:2015qla} and those predictions in Refs.~\cite{Zhang:2018ubo,Godfrey:2016nwn,Asghar:2018tha,Sun:2014wea}. The decay width is nearly saturated by
the $B^*\pi$ channel. If we taking the mixing
angle around the value obtained in the heavy-quark
symmetry limit, i.e. $\theta_{1P}=-(55\pm 5)^{\circ}$~\cite{Isgur:1989vq,Isgur:1989ed,Close:2005se,Zhong:2008kd,Matsuki:2010zy,Barnes:2002mu}, the decay width is predicted to be in the range of $\Gamma=(24.5\pm 2.5)$ MeV, which seems to
be more comparable with the LHCb observations~\cite{Aaij:2015qla}.
We also study the radiative decay processes
of $B_1(5721)\to B^{*}\gamma,B\gamma$, their partial decay widths
are predicted to be
\begin{eqnarray}\label{ra}
\Gamma [B_1(5721)^{0}\to B^{*0}\gamma/B^{0}\gamma]&=&66/24 \ \mathrm{keV},\\
\Gamma [B_1(5721)^{+}\to B^{*+}\gamma/B^{+}\gamma]&=&206/69 \ \mathrm{keV}.
\end{eqnarray}
Our predictions are comparable with those in Ref.~\cite{Orsland:1998de},
however, most of our predictions are notably smaller than the predictions in Refs.~\cite{Lu:2016bbk,Godfrey:2016nwn,Asghar:2018tha}.
The radiative decay modes $B^*\gamma$ and $B\gamma$ of
$B_1(5721)$ may be observed in future experiments since their
branching fractions can reach up to the order of $O(10^{-3})$.

In the $B$-meson family, the mass for the other mixed
state $B_{1}(1P_1)$ is about 40 MeV lower
than that of $B_1(5721)$ according to our potential model calculations,
which is consistent with the prediction in Ref.~\cite{DiPierro:2001dwf}.
A slightly smaller mass splitting, $\sim (10-30)$ MeV,
between $B_1(5721)$ and $B_{1}(1P_1)$ is given in Refs.~\cite{Kher:2017mky,Ebert:2009ua,Lu:2016bbk,
Godfrey:2016nwn,Asghar:2018tha}. Thus, the mass of $B_{1}(1P_1)$ might be in
the range of $(5700\pm 15)$ MeV. Considering the mass uncertainties,
with the mixing angle $\theta_{1P}=-35.2^{\circ}$ we plot the decay width
of $B_{1}(1P_1)$ as a function of its mass in Fig.~\ref{b1p}. It is found
that $B_{1}(1P_1)$ is a broad state with a width of $\Gamma\simeq (155\pm 10)$ MeV.
The $B^*\pi$ channel is the only OZI-allowed two body strong decay channel.
The $B_J(5732)$ listed in the RPP~\cite{Zyla:2020zbs} is a good candidate for the $B_{1}(1P_1)$.
With this assignment, both the measured mass $M_{exp}=5698$ MeV and width
$\Gamma_{exp}=(128\pm 18)$ MeV for the $B_J(5732)$ are in
good agreement with the quark model predictions. However, in Ref.~\cite{Shah:2016mgq} the $B_J(5732)$
is assigned to the $B(1^3P_0)$ state according to the mass spectrum analysis.

As the mixed state $|1P'_1\rangle$, the mass of $B_{s1}(5830)$ is
consistent with our fitted value $M=5821$ MeV and
other determinations in the literature (See Table~\ref{massBs}).
It dominantly decays into the $B^*K$ channel.
Taking the mixing angle $\theta_{1P}=-39.6^{\circ}$ determined
by our potential model, we find that the theoretical width,
$\Gamma\simeq 7.6$ MeV, is too large to be
comparable with the measured one $\Gamma_{exp}\simeq (0.5\pm 0.3\pm 0.3)$ MeV.
However, if we taking the mixing angle around the value obtained in the heavy-quark
symmetry limit, i.e. $\theta_{1P}=-(55\pm 5)^{\circ}$,
as that adopted in Ref.~\cite{Zhong:2008kd}, the decay width is predicted
to be in the range of $\Gamma=0.08-0.8$ MeV, which is comparable with the data.
It indicates that the mixing angle between $1^1P_1$ and $1^3P_1$
may be close to the value $\theta_{1P}=-55^{\circ}$ obtained in the heavy-quark
symmetry limit. The $B_{s1}(5830)$ has large decay rates into $B_s^{*}\gamma$
and $B_s\gamma$ channels. Their partial decay widths are predicted to be
\begin{eqnarray}\label{ra}
\Gamma [B_{s1}(5830)\to B^{*}_s\gamma]&=&53 \ \mathrm{keV},\\
\Gamma [B_{s1}(5830)\to B_s\gamma]&=&27 \ \mathrm{keV}.
\end{eqnarray}
The branching fractions of these radiative decays may reach up to
$O(10^{-2})$. Our predictions are comparable with those predicted in
Refs.~\cite{Lu:2016bbk,Godfrey:2016nwn,Orsland:1998de}. The $B_s^{*}\gamma$
and $B_s\gamma$ decay channels of $B_{s1}(5830)$ may have good potentials to
be observed in future experiments.

In the $B_s$ meson sector, the mass of the other
mixed state $B_{s1}(1P_1)$ is predicted to be about $2-40$ MeV lower than that
of $B_{s1}(5830)$ in various quark models~\cite{Godfrey:2016nwn,Kher:2017mky,Ebert:2009ua,Zeng:1994vj,
Lu:2016bbk,Asghar:2018tha,DiPierro:2001dwf}. Thus, the mass
of $B_{s1}(1P_1)$ is estimated to be $\sim (5808\pm 19)$ MeV, which is just around
the $B^{*+}K^-$ and $B^{*0}K^0$ mass thresholds. From Fig.~\ref{b1p}, it is seen that
the strong decay properties of $B_{s1}(1P_1)$ are very sensitive to the mass threshold.
There are three cases to be considered.
(i) If the mass of $B_{s1}(1P_1)$ is below
the $B^{*+}K^-$ mass threshold 5818 MeV, the radiative decay modes
$B^{*}\gamma$ and $B\gamma$ may play crucial roles in the decays.
The with partial decay widths are estimated to be
\begin{eqnarray}\label{ra}
\Gamma [B_{s1}(1P_1)\to B^{*}_s\gamma]&\simeq &60 \ \mathrm{keV},\\
\Gamma [B_{s1}(1P_1)\to B_s\gamma]&\simeq &40 \ \mathrm{keV}.
\end{eqnarray}
Then, the $B_{s1}(1P_1)$ should has a very narrow width of $\Gamma\sim \mathcal{O}(100)$ keV.
(ii) If the mass of $B_{s1}(1P)$ lies between the $B^{*+}K^-$ mass threshold 5818 MeV
and the $B^{*0}K^0$ mass threshold 5822 MeV, the $B_{s1}(1P_1)$ dominantly decays
into $B^{*0}K^0$ mode, and has a narrow width of $\Gamma\simeq (20\pm 15)$ MeV.
(iii) If the mass of $B_{s1}(1P_1)$ is above the $B^{*0}K^0$ mass threshold 5822 MeV,
the $B_{s1}(1P_1)$ dominantly decays into $B^{*0}K^0$
and $B^{*+}K^-$ channels, and has a relatively broad width of $\Gamma\simeq (70\pm30)$ MeV.
It should be mentioned that the OPAL Collaboration observed some signals of
a resonance denoted by $B_{sJ}(5850)$ with a mass of $M_{exp}=(5853\pm 15)$ MeV
and a width of $\Gamma\simeq (47\pm 22)$ MeV many year ago~\cite{Akers:1994fz}. As a candidate
of $B_{s1}(1P_1)$, both the measured mass and width of $B_{sJ}(5850)$ are consistent
with the predictions. To confirm the $B_{sJ}(5850)$ resonance and established
the $B_{s1}(1P_1)$ state, more observations of the $B^{*0}K^0$ and $B^{*+}K^-$
final states are suggested to be carried out in future experiments.

As a whole the high mass mixed state $|1P'_1\rangle$ via the $1^1P_1$-$1^3P_1$ mixing
in the $B$ and $B_s$ families have been well-established, they correspond to two narrow states $B_1(5721)$
and $B_{s1}(5830)$ observed in experiments. Some evidence for the low mass
mixed states $|1P_1\rangle$ with broad widths predicted in theory may
have been observed in experiments. The $B_{J}(5732)$ and $B_{sJ}(5850)$
resonances listed by the PDG~\cite{Zyla:2020zbs} may good candidates for the missing $|1P_1\rangle$
in the $B$ and $B_s$ families, respectively.

\begin{table*}[htb]%£¨except 11S0£©
\caption{Partial and total decay widths (MeV) for the $2S$-wave $B$ and $B_s$ mesons. }\label{decayB2s}
\begin{tabular}{cccccccccccccccccccccc}  \midrule[1.0pt]\midrule[1.0pt]
 {\multirow{2}{*}{$n^{2S+1}L_J$}} ~~~~& \multirow{2}{*}{State}~~~~& \multirow{2}{*}{Channel}~~~~& \multicolumn{1}{c} {\underline{~~~~$\Gamma_{th}$ (MeV)~~~~}}
 ~~~~& \multirow{2}{*}{State}~~~~& \multirow{2}{*}{Channel}~~~~& \multirow{2}{*}{$\Gamma_{th}$ (MeV)}~~~~
 \\
~~~~&~~~~&~~~~&($b\bar{u}$~/~$b\bar{d}$)~~~~&~~~~&~~~~&~~~~&~~~~\\
\midrule[1.0pt]	                                                   % ME               zhong                          liu           lv             exp
$2^1S_0$     ~~~~&     $B_0(5876)$   ~~~~&   $B^*\pi  $          ~~~~&  59                                   &$B_{s0}(5944)$ &   $B^*K$               ~~~~&  55          \\
             ~~~~&                   ~~~~&   $B^*\eta $          ~~~~&  0.05                                 & ~~~~&             $B_s^*\eta$          ~~~~&  $\cdots$    \\
             ~~~~&                   ~~~~&   $B(1^3P_0)\pi$      ~~~~&  4.8                                  & ~~~~&             $B^*_s\gamma$        ~~~~&  0.1$\times 10^{-3}$         \\
             ~~~~&                   ~~~~&   $B^*\gamma$         ~~~~&  0.006/0.85$\times 10^{-3}$                          & ~~~~&             $B_s(1P)\gamma$      ~~~~&  0.009          \\
             ~~~~&                   ~~~~&   $B(1P)\gamma$       ~~~~&  0.1/0.034                            & ~~~~&             $B_{s1}(5830)\gamma$ ~~~~& 4.9$\times 10^{-3}$         \\
             ~~~~&                   ~~~~&   $B_1(5721)\gamma$   ~~~~&  0.026/8.8$\times 10^{-3}$            & ~~~~&              Total               ~~~~&  55          \\
             ~~~~&                   ~~~~&   Total               ~~~~&  63                                   & ~~~~&                                  ~~~~&              \\
$2^3S_1$     ~~~~&    $B_1(5899)$    ~~~~&   $B\pi  $            ~~~~&  5.1                                  &$B_{s1}(5966)$&    $BK$                 ~~~~&  15         \\
             ~~~~&                   ~~~~&   $B\eta $            ~~~~&  0.8                                  & ~~~~&             $B_s\eta$            ~~~~&  0.1        \\
             ~~~~&                   ~~~~&   $B_sK $             ~~~~&  0.02                                 & ~~~~&             $B^*K$               ~~~~&  32         \\
             ~~~~&                   ~~~~&   $B^*\pi $           ~~~~&  22                                   & ~~~~&             $B_s^*\eta$          ~~~~&  0.01        \\
             ~~~~&                   ~~~~&   $B^*\eta $          ~~~~&  0.6                                  & ~~~~&             $B_s  \gamma$        ~~~~&  3.6$\times 10^{-3}$        \\
             ~~~~&                   ~~~~&   $B_2(5747)\pi$      ~~~~&  $<0.01$                              & ~~~~&             $B_s(2^1S_0)\gamma$  ~~~~& 1.0$\times 10^{-6}$      \\
             ~~~~&                   ~~~~&   $B(1P)\pi$          ~~~~&  11                                  & ~~~~&             $B_{s2}(5840)\gamma$ ~~~~& 9$\times 10^{-3}$    \\
             ~~~~&                   ~~~~&   $B_1(5721)\pi$      ~~~~&  1                                    & ~~~~&             $B_s(1P)\gamma$      ~~~~&  2.2$\times 10^{-3}$        \\
             ~~~~&                   ~~~~&   $B  \gamma$         ~~~~&  0.029/0.008                          & ~~~~&             $B_{s1}(5830)\gamma$ ~~~~& 2.7$\times 10^{-3}$        \\
             ~~~~&                   ~~~~&   $B(2^1S_0)\gamma$   ~~~~&  6$\times 10^{-5}$/2$\times 10^{-5}$  & ~~~~&             $B(1^3P_0)\gamma$    ~~~~&1.4$\times 10^{-3}$        \\
             ~~~~&                   ~~~~&   $B_2(5747)\gamma$   ~~~~&  0.06/0.019                         &  ~~~~&             Total                ~~~~&  47         \\
             ~~~~&                   ~~~~&   $B(1P)\gamma$       ~~~~&  0.015/5.2$\times 10^{-3}$            & ~~~~&                                  ~~~~&            \\
             ~~~~&                   ~~~~&   $B_1(5721)\gamma$   ~~~~& 0.018/6.2$\times 10^{-3}$             & ~~~~&                                  ~~~~&            \\
             ~~~~&                   ~~~~&   $B(1^3P_0)\gamma$   ~~~~&  0.004/1.6$\times 10^{-3}$            & ~~~~&                                  ~~~~&            \\
             ~~~~&                   ~~~~&   Total               ~~~~&  41                                   & ~~~~&                                  ~~~~&            \\

\midrule[1.0pt]\midrule[1.0pt]
\end{tabular}
\end{table*}

\subsection{$2S$-wave states}

\subsubsection{$2^1S_0$ states}

The $2^1S_0$ states in the $B$ and $B_s$ families are still not established.
In the $B$ meson sector, our predicted mass for the $B(2^1S_0)$ state
is $M=5876$ MeV, which is in agreement with the predictions in Refs.~\cite{Ebert:2009ua,Godfrey:2016nwn,
Asghar:2018tha,DiPierro:2001dwf}. Taking the mass $M=5876$ MeV, we calculate the
strong and radiative decay properties of $B(2^1S_0)$, our results are listed
in Table~\ref{decayB2s}. The decays of $B(2^1S_0)$ are governed by the
$B^*\pi$ mode with a fairly large branching fraction $\sim 94\%$.
The width of $B(2^1S_0)$ is predicted to be $\Gamma\simeq 63$ MeV,
which is about a factor $1.5-4$ larger than the predictions in Refs.~\cite{Xiao:2014ura,DiPierro:2001dwf,Sun:2014wea,Asghar:2018tha},
while about a factor $1.5-2$ smaller than the predictions in Refs.~\cite{Lu:2016bbk,Godfrey:2016nwn,Yu:2019iwm,Ferretti:2015rsa}.

In 2015, the LHCb Collaboration observed two new resonances
$B_{J}(5840)^{0,+}$~\cite{Aaij:2015qla}. Considering $B_{J}(5840)^{0,+}$ as unnatural parity states,
the relatively accurate measurements of the mass and width
for the neutral one are $M_{exp}=(5863\pm9)$ MeV and $\Gamma_{exp}=(127\pm 51)$ MeV,
respectively~\cite{Aaij:2015qla}. In this case, signals of $B_{J}(5840)^{0,+}$
should come from the $B^*\pi$ decay mode other than $B\pi$.
In Refs.~\cite{Godfrey:2016nwn,Jia:2018vwl,Gupta:2017bcm,Lu:2016bbk,Asghar:2018tha},
the $B_{J}(5840)^{0,+}$ resonances were suggested to be the $B(2^1S_0)$ assignment.
If $B_{J}(5840)^{0,+}$ have an unnatural parity indeed, they strongly favor the
$B(2^1S_0)$ assignment. The measured mass and width together with the decay modes
of $B_{J}(5840)$ are consistent with the our theoretical predictions.
However, a natural parity for $B_{J}(5840)^{0,+}$ is also possible according
to the LHCb analysis (see Case B listed in Table \ref{Excitation}).
The $B_{J}(5840)^{0,+}$ may possibly decay into both the $B^*\pi$ and $B\pi$ channels.
If the $B\pi$ decay mode is confirmed in future experiments, the $B_{J}(5840)$
resonance should be other assignments since the $B\pi$ mode is forbidden
for the $B(2^1S_0)$ state.

In the $B_s$ meson sector, our predicted mass for the $B_s(2^1S_0)$ state
is $M=5944$ MeV, which is comparable with the other quark model predictions~\cite{Kher:2017mky,Ebert:2009ua,Zeng:1994vj,Lu:2016bbk,Asghar:2018tha,Godfrey:2016nwn,DiPierro:2001dwf}.
Taking $M=5944$ MeV, we calculate the
strong and radiative decay properties of $B_s(2^1S_0)$, our results are listed in Table~\ref{decayB2s}.
The $B^*K$ channel is the only OZI-allowed two body strong decay channel for $B_s(2^1S_0)$.
Its decay width is predicted to be $\Gamma\simeq 55$ MeV, which is comparable with those predictions in Refs.~\cite{Xiao:2014ura,Godfrey:2016nwn,Asghar:2018tha,Sun:2014wea,Ferretti:2015rsa}.
The $B_s(2^1S_0)$ state should have large potentials to be seen in the $B^{*+}K^-$ channel
since it has a fairly narrow width.

\subsubsection{$2^3S_1$ states}

In the $B$ meson sector, our predicted mass for $B(2^3S_1)$ is $M=5899$ MeV,
which is in good agreement with the predictions
in Refs.~\cite{Ebert:2009ua,Asghar:2018tha}. Taking the mass $M=5899$ MeV, we calculate the
strong and radiative decay properties of $B(2^3S_1)$, our results are listed
in Table~\ref{decayB2s}. It is found that the $B(2^3S_1)$ state is a fairly
narrow state with a width of $\Gamma \simeq 40$ MeV, which is consistent
with the prediction in Refs.~\cite{Xiao:2014ura,Sun:2014wea}, however, a factor $\sim 2-4$ smaller than
the predictions in Refs.~\cite{Ferretti:2015rsa,Godfrey:2016nwn,Yu:2019iwm,Lu:2016bbk}. This state dominantly decays
into the $B^*\pi$ channel with a branching fraction $\sim 54\%$. The partial width ratio
between the two typical channels $B\pi$ and $B^*\pi$ is predicted to be
\begin{eqnarray}\label{ra}
\frac{\Gamma[B\pi]}{\Gamma[B^*\pi]}\simeq 0.23,
\end{eqnarray}
which can be used to identify the $B(2^3S_1)$ from its possible
candidates observed in future experiments.

It should be mentioned that the $B_{J}(5840)$ resonance
may be a candidate of the $B(2^3S_1)$ state as suggested in Refs.~\cite{Godfrey:2019cmi,Yu:2019iwm}.
As this assignment, both the mass and typical decay modes $B^*\pi$ and $B\pi$
predicted in theory are consistent with the LHCb observations~\cite{Aaij:2015qla}.
Our predicted width $\Gamma \simeq 41$ MeV is close to the lower limit
of the measured width $\Gamma_{exp}=(107\pm 54)$ MeV assuming a natural parity for $B_{J}(5840)^0$~\cite{Aaij:2015qla}.
Furthermore, the $B_{J}(5970)$ resonance was also
suggested to be the $B(2^3S_1)$ state in the literature~\cite{Xu:2014mka,Jia:2018vwl,
Sun:2014wea,Lu:2016bbk,Godfrey:2016nwn,Ferretti:2015rsa}. The $B_{J}(5970)^{0,+}$ resonances were first observed by the CDF Collaboration
in the $B\pi$ final states in 2013~\cite{Aaltonen:2013atp}, and confirmed by the LHCb
Collaboration two years later~\cite{Aaij:2015qla}. The central value of the measured width is
$\Gamma_{exp}\simeq60-70$ MeV with large uncertainties (see Table~\ref{Excitation}).
In the LHCb observations, the $B^*\pi$ decay mode has been seen,
while the $B\pi$ mode may be possibly seen~\cite{Zyla:2020zbs}.
If assigning $B_J(5970)$ resonance to $B(2^3S_1)$, our predicted mass, $M\simeq 5899$ MeV, is
about 70 MeV larger than the observation, while the predicted width, $\Gamma\simeq 54$ MeV,
is consistent with the data. As a conclusion, we cannot exclude
the possibilities of the $B_{J}(5840)$ and $B_J(5970)$ resonances as candidates of the $B(2^3S_1)$ state
based on the present experimental information.

In the $B_s$ meson sector, our predicted mass for the $B_s(2^3S_1)$
state is $M=5966$ MeV, which is in good agreement with the predictions
in Refs.~\cite{Zeng:1994vj,Ebert:2009ua,Asghar:2018tha}. Taking $M=5966$ MeV, we calculate the
strong and radiative decay properties of $B_s(2^3S_1)$, our results are listed
in Table~\ref{decayB2s}. It is found that the $B_s(2^3S_1)$ state is also a
narrow state with a width of $\Gamma \simeq 50$ MeV, which is consistent
with the predictions in Refs.~\cite{Xiao:2014ura,Sun:2014wea}, however, a factor $\sim 2-4$ smaller than
the predictions in Refs.~\cite{Godfrey:2016nwn,Lu:2016bbk,Ferretti:2015rsa}. This state mainly decays
into the $B^*K$ and $BK$ channel with large branching fractions $\sim 68\%$ and $\sim 32\%$,
respectively. The partial width ratio between the two main channels $BK$ and $B^*K$ is predicted to be
\begin{eqnarray}\label{ra}
\frac{\Gamma[BK]}{\Gamma[B^*K]}\simeq 0.47,
\end{eqnarray}
which may be an important criterion for establishing the $B_s(2^3S_1)$.

There may exist a strong configuration mixing
between the $2^3S_1$ and $1^3D_1$ states, which is to be
discussed in the last part of this section.

\begin{table*}[htb]%£¨except 11S0£©
\caption{Partial and total decay widths (MeV) for the $1D$-wave $B$ and $B_s$ mesons.}\label{decay2d1}
\begin{tabular}{cccccccccccccccccccccc}  \midrule[1.0pt]\midrule[1.0pt]
 {\multirow{2}{*}{$n^{2S+1}L_J$}} ~~~~~~~~~~~~&  \multicolumn{1}{c} {\underline{Observed State}}~~~~& \multirow{2}{*}{Channel}~~~~& \multicolumn{1}{c} {\underline{~~~~$\Gamma_{th}$ (MeV)~~~~}}
~~~~&   ~~~~~~~~~~~~&  \multicolumn{1}{c} {\underline{Observed State}}~~~~& \multirow{2}{*}{Channel}~~~~& \multirow{2}{*}{$\Gamma_{th}$ (MeV)}~~~~
 \\
~~~~~~~~~~~~&$B$ meson~~~~&~~~~~~~~~~~~&($b\bar{u}$~/~$b\bar{d}$)~~~~&~~~~&$B_s$ meson~~~~&~~~~&~~~~&~~~~\\
\midrule[1.0pt]		                                                   % ME               zhong                          liu           lv             exp
$1^3D_3$     ~~~~~~~~~~~~&$B^*_3(5979)  $    ~~~~&   $B\pi  $            ~~~~&  17         ~~~~&~~~~&$B^*_{s3}(6067)  $ ~~~~&   $BK$                ~~~~&  7.0        \\
             ~~~~~~~~~~~~&                   ~~~~&   $B\eta $            ~~~~&  0.2 ~~~~&~~~~&  ~~~~&   $B_s\eta$           ~~~~&  0.1        \\
             ~~~~~~~~~~~~&                   ~~~~&   $B_sK $             ~~~~&  0.1 ~~~~&~~~~&  ~~~~&   $B^*K$              ~~~~&  5.8        \\
             ~~~~~~~~~~~~&                   ~~~~&   $B^*\pi $           ~~~~&  17  ~~~~&~~~~&  ~~~~&   $B_s^*\eta$         ~~~~&  0.03       \\
             ~~~~~~~~~~~~&                   ~~~~&   $B^*\eta $          ~~~~&  0.1~~~~&~~~~&   ~~~~&   $B_{s2}(5840)\gamma$ ~~~~&  0.038         \\
             ~~~~~~~~~~~~&                   ~~~~&   $B_s^*K $           ~~~~&  0.01 ~~~~&~~~~& ~~~~&   $B_{s1}(5830)\gamma$     ~~~~&  2.0$\times 10^{-4}$        \\
             ~~~~~~~~~~~~&                   ~~~~&   $B_2(5747)\pi $     ~~~~&  0.9 ~~~~&~~~~&  ~~~~&   $B_s(1P_1)\gamma$    ~~~~&  2.0$\times 10^{-4}$        \\
             ~~~~~~~~~~~~&                   ~~~~&   $B_1(5721)\pi $     ~~~~&  0.01~~~~&~~~~&  ~~~~&   $B_s(1^3P_0)\gamma$ ~~~~&  1.0$\times 10^{-4}$       \\
             ~~~~~~~~~~~~&                   ~~~~&   $B(1P_1)\pi $       ~~~~&  3.6   ~~~~&~~~~& ~~~~&   Total               ~~~~&  13         \\
             ~~~~~~~~~~~~&                   ~~~~&   $B_2(5747)\gamma$   ~~~~&  0.125/0.043~~~~&~~~~& ~~~~&                ~~~~&             \\
             ~~~~~~~~~~~~&                   ~~~~&   $B_1(5721)\gamma$   ~~~~&  1.9$\times 10^{-3}$/0.5$\times 10^{-3}$    ~~~~&~~~~&                   ~~~~&                       ~~~~&             \\
             ~~~~~~~~~~~~&                   ~~~~&   $B(1P_1)\gamma$     ~~~~&  1.9$\times 10^{-3}$/0.5$\times 10^{-3}$    ~~~~&~~~~&                   ~~~~&                       ~~~~&             \\
             ~~~~~~~~~~~~&                   ~~~~&   $B(1^3P_0)\gamma$   ~~~~&  0.7$\times 10^{-3}$/0.2$\times 10^{-3}$    ~~~~&~~~~&                   ~~~~&                       ~~~~&             \\
             ~~~~~~~~~~~~&                   ~~~~&   Total               ~~~~&  39         ~~~~&~~~~&                   ~~~~&                       ~~~~&             \\
\\
$1^3D_1$     ~~~~~~~~~~~~&$B^*_1(6056)  $    ~~~~&   $B\pi  $            ~~~~&  92         ~~~~&~~~~&$B^*_{s1}(6101)$   ~~~~&   $BK$                ~~~~& 87          \\
             ~~~~~~~~~~~~&                   ~~~~&   $B\eta $            ~~~~&  14         ~~~~&~~~~&                   ~~~~&   $B_s\eta$           ~~~~& 7.3         \\
             ~~~~~~~~~~~~&                   ~~~~&   $B_sK $             ~~~~&  21         ~~~~&~~~~&                   ~~~~&   $B^*K$              ~~~~& 37          \\
             ~~~~~~~~~~~~&                   ~~~~&   $B^*\pi $           ~~~~&  41         ~~~~&~~~~&                   ~~~~&   $B_s^*\eta$         ~~~~& 2.6         \\
             ~~~~~~~~~~~~&                   ~~~~&   $B^*\eta $          ~~~~&  5.3   ~~~~&~~~~&    ~~~~&   $B_{s2}(5840)\gamma$ ~~~~& 3.9$\times 10^{-3}$         \\
             ~~~~~~~~~~~~&                   ~~~~&   $B_s^*K $           ~~~~&  7.3   ~~~~&~~~~&    ~~~~&   $B_s(1P_1)\gamma$    ~~~~& 0.017          \\
             ~~~~~~~~~~~~&                   ~~~~&   $B_2(5747)\pi $     ~~~~&  2.1   ~~~~&~~~~&    ~~~~&   $B_{s1}(5830)\gamma$    ~~~~& 0.023          \\
             ~~~~~~~~~~~~&                   ~~~~&   $B_1(5721)\pi $     ~~~~&  148  ~~~~&~~~~&    ~~~~&   $B_s(1^3P_0)\gamma$ ~~~~& 0.044          \\
             ~~~~~~~~~~~~&                   ~~~~&   $B(1P)\pi $         ~~~~&  20        ~~~~&~~~~&                   ~~~~&   Total               ~~~~& 133         \\
             ~~~~~~~~~~~~&                   ~~~~&   $B_2(5747)\gamma$   ~~~~&  0.035/0.01      ~~~~&~~~~&                   ~~~~&                       ~~~~&             \\
             ~~~~~~~~~~~~&                   ~~~~&   $B(1P)\gamma$       ~~~~&  0.122/0.039     ~~~~&~~~~&                   ~~~~&                       ~~~~&             \\
             ~~~~~~~~~~~~&                   ~~~~&   $B_1(5721)\gamma$   ~~~~&  0.178 /0.056     ~~~~&~~~~&                   ~~~~&                       ~~~~&             \\
             ~~~~~~~~~~~~&                   ~~~~&   $B(1^3P_0)\gamma$   ~~~~&  0.201/0.066     ~~~~&~~~~&                   ~~~~&                       ~~~~&             \\
             ~~~~~~~~~~~~&                   ~~~~&   Total               ~~~~&  350        ~~~~&~~~~&                   ~~~~&                       ~~~~&             \\
\\
$1D'_2$      ~~~~~~~~~~~~&$B_2(6067)  $      ~~~~&   $B^*\pi $           ~~~~&  66         ~~~~&~~~~&$B_{s2}(6113)  $   ~~~~&   $B^*K$              ~~~~&  22         \\
             ~~~~~~~~~~~~&                   ~~~~&   $B^*\eta $          ~~~~&  1.7        ~~~~&~~~~&                   ~~~~&   $B_S^*\eta$         ~~~~&  0.5        \\
             ~~~~~~~~~~~~&                   ~~~~&   $B_s^*K $           ~~~~&  1.4   ~~~~&~~~~& ~~~~&   $B_{s2}(5840)\gamma$ ~~~~&  0.016         \\
             ~~~~~~~~~~~~&                   ~~~~&   $B(1^3P_0)\pi $     ~~~~&  3.5   ~~~~&~~~~& ~~~~&   $B_s(1P)\gamma$    ~~~~&  1.3$\times 10^{-3}$    \\
             ~~~~~~~~~~~~&                   ~~~~&   $B_2(5747)\pi $     ~~~~&  6.2   ~~~~&~~~~& ~~~~&   $B_{s1}(5830)\gamma$     ~~~~&  0.064         \\
             ~~~~~~~~~~~~&                   ~~~~&   $B_1(5721)\pi $     ~~~~&  10  ~~~&~~~~& ~~~~&   $B_s(1^3P_0)\gamma$ ~~~~&  3.0$\times 10^{-4}$        \\
             ~~~~~~~~~~~~&                   ~~~~&   $B(1P )\pi $        ~~~~&  0.4        ~~~~&~~~~&                   ~~~~&   Total               ~~~~&  23         \\
             ~~~~~~~~~~~~&                   ~~~~&   $B_2(5747)\gamma$   ~~~~&  0.106/0.033     ~~~~&~~~~&                   ~~~~&                       ~~~~&             \\
             ~~~~~~~~~~~~&                   ~~~~&   $B(1P)\gamma$       ~~~~&  0.022/6.5$\times 10^{-3}$     ~~~~&~~~~&                   ~~~~&                       ~~~~&             \\
             ~~~~~~~~~~~~&                   ~~~~&   $B_1(5721)\gamma$   ~~~~&  0.302/0.104     ~~~~&~~~~&                   ~~~~&                       ~~~~&             \\
             ~~~~~~~~~~~~&                   ~~~~&   $B(1^3P_0)\gamma$   ~~~~&  3.3$\times 10^{-3}$/1.0$\times 10^{-3}$    ~~~~&~~~~&                   ~~~~&                       ~~~~&             \\
             ~~~~~~~~~~~~&                   ~~~~&   Total               ~~~~&  90        ~~~~&~~~~&                   ~~~~&                       ~~~~&             \\
\\
$1D_2$       ~~~~~~~~~~~~&$B_2(5973)  $      ~~~~&   $B^*\pi $           ~~~~&  92         ~~~~&~~~~&$B_{s2}(6061)  $   ~~~~&   $B^*K$              ~~~~&  85         \\
             ~~~~~~~~~~~~&                   ~~~~&   $B^*\eta $          ~~~~&  6.7        ~~~~&~~~~&                   ~~~~&   $B_S^*\eta$         ~~~~&  4.2        \\
             ~~~~~~~~~~~~&                   ~~~~&   $B_s^*K $           ~~~~&  5.2 ~~~~&~~~~&   ~~~~&   $B_{s2}(5840)\gamma$ ~~~~&  6.0$\times 10^{-3}$           \\
             ~~~~~~~~~~~~&                   ~~~~&   $B(1^3P_0)\pi $     ~~~~&  0.002~~~~&~~~~&   ~~~~&   $B_s(1P)\gamma$    ~~~~&  0.045         \\
             ~~~~~~~~~~~~&                   ~~~~&   $B_2(5747)\pi $     ~~~~&  64  ~~~~&~~~~&   ~~~~&   $B_{s1}(5830)\gamma$     ~~~~& 0.5$\times 10^{-3}$         \\
             ~~~~~~~~~~~~&                   ~~~~&   $B(1P)\pi $         ~~~~&  0.4 ~~~~&~~~~&   ~~~~&   $B_s(1^3P_0)\gamma$ ~~~~&  0.1$\times 10^{-3}$        \\
             ~~~~~~~~~~~~&                   ~~~~&   $B_1(5721)\pi $     ~~~~&  0.04~~~~&~~~~&    ~~~~&   Total               ~~~~&  90         \\
             ~~~~~~~~~~~~&                   ~~~~&   $B_2(5747)\gamma$   ~~~~&  0.026/0.008      ~~~~&~~~~&                   ~~~~&                       ~~~~&  \\
             ~~~~~~~~~~~~&                   ~~~~&   $B(1P)\gamma$       ~~~~&  0.223/0.076    ~~~~&~~~~&                   ~~~~&                       ~~~~&             \\
             ~~~~~~~~~~~~&                   ~~~~&   $B_1(5721) \gamma$  ~~~~&  1.2$\times 10^{-3}$/0.3$\times 10^{-3}$~~~~&~~~~&                   ~~~~&                       ~~~~&             \\
             ~~~~~~~~~~~~&                   ~~~~&   $B(1^3P_0)\gamma$   ~~~~&  0.5$\times 10^{-3}$/0.2$\times 10^{-3}$    ~~~~&~~~~&                   ~~~~&                       ~~~~&             \\
             ~~~~~~~~~~~~&                   ~~~~&   Total               ~~~~&  169        ~~~~&~~~~&                   ~~~~&                       ~~~~&             \\

\midrule[1.0pt]\midrule[1.0pt]
\end{tabular}
\end{table*}

\subsection{$1D$-wave states}

Some evidence of the $1D$-wave $B$ and $B_s$ states may have been observed in experiments.
The $B(5970)^{0,+}$ together with the new resonances $B_{sJ}(6064)$ and $B_{sJ}(6114)$
observed at LHCb may be good candidates of the $1D$-wave states according to
the mass spectrum predictions in various quark models.

\subsubsection{$1^3D_3$ states}

In Ref.~\cite{Xiao:2014ura}, by analyzing the decay properties within the chiral quark model
our group found that the $B_J(5970)$ is most likely to be the $1^3D_3$ assignment in the $B$-meson family.
The $B_{J}(5970)$ as a candidate of $B(1^3D_3)$ is also suggested in
Refs.~\cite{Yu:2019iwm,Lu:2016bbk}. In present work, we restudy the $B_J(5970)$ by combining the decay properties
with the mass spectrum. It is found that as the $B(1^3D_3)$ assignment the mass of $B(5970)$
can be well explained with the potential model. Our predicted mass
$M=5979$ MeV is in good agreement with the observed value $M_{exp}=(5971\pm 5)$ MeV for
the neutral state $B(5970)^0$~\cite{Zyla:2020zbs}.
By using the wave function of $B(1^3D_3)$
calculated from the potential model, we further study the decay
properties, our results are listed in Table~\ref{decay2d1}.
It is found that the predicted decay width,
$\Gamma\simeq 39$ MeV, is also consistent the measured width
$\Gamma_{exp}\simeq (56\pm 16)$ MeV of $B(5970)^0$ by assuming $P=(-1)^J$ and using three
relativistic Breit-Wigner functions in the fit for
mass difference at LHCb~\cite{Aaij:2015qla}. The partial
width ratio between $B\pi$ and $B^*\pi$ is predicted to be
\begin{eqnarray}\label{ra}
\frac{\Gamma[B\pi]}{\Gamma[B^*\pi]}\simeq 1.0,
\end{eqnarray}
which is waiting to be tested in future experiments.
If the $B_{J}(5970)^{0,+}$ resonances correspond to the $B(1^3D_3)$ assignment indeed, the charged
state $B_{J}(5970)^+$ should have a large radiative decay rate into
$B_2^*(5747)^+\gamma$, the partial width and branching fraction
are predicted to be
\begin{eqnarray}\label{5747}
\Gamma [B_{J}(5970)^+\to B_2^*(5747)^+\gamma] &\simeq & 125\ \mathrm{keV},\\
Br[B_{J}(5970)^+\to B_2^*(5747)^+\gamma]& \simeq & 3\times 10^{-3}.
\end{eqnarray}
The radiative decay mode $B_2^*(5747)^+\gamma$ may be observed in future experiments.

In Ref.~\cite{Xiao:2014ura}, considering the $B_J(5970)$ as the $B(1^3D_3)$ assignment, our group further predicted
that the mass and width of the $B_s(1^3D_3)$ state, as a flavor partner of $B_J(5970)$,
might be $M\simeq 6.07$ GeV and $\Gamma\simeq 30$ MeV, respectively.
The predicted mass of $B_s(1^3D_3)$ is consistent with the those predicted in Refs.~\cite{Zeng:1994vj,Lu:2016bbk,Asghar:2018tha}, while
a relatively narrow width is also predicted by other works~\cite{Asghar:2018tha,Xu:2014mka,Sun:2014wea,Lu:2016bbk,Godfrey:2016nwn,Ferretti:2015rsa}.
It is interestingly found that the new bottom-strange structure
$B_{sJ}(6064)$ with a mass of $M_{exp}=(6063.5\pm 2.0)$ MeV and a very narrow width
of $\Gamma_{exp}=(26\pm 8)$ MeV observed at LHCb~\cite{Aaij:2020hcw} is consistent with the predictions.
In present work, from the aspects of both mass spectrum and decay
properties we further discuss the possibility of the $B_{sJ}(6064)$
structure as the $B_s(1^3D_3)$ assignment in the $B_s$-meson family.
With this assignment, it is found that the measured mass for
the $B_{sJ}(6064)$ structure is consistent with the theoretical mass
$M=6067$ MeV. Furthermore, the narrow width of $B_{sJ}(6064)$ can also be explained
within our chiral quark model. From our predicted decay properties
listed in Table~\ref{decay2d1}, it is found that the theoretical
width $\Gamma\simeq 13$ MeV is close to lower limit of the measured
value $\Gamma_{exp}=(26\pm 8)$ MeV from LHCb~\cite{Aaij:2020hcw}. The partial
width ratio between $BK$ and $B^*K$ is predicted to be
\begin{eqnarray}\label{ra}
\frac{\Gamma[BK]}{\Gamma[B^*K]}\simeq 1.2.
\end{eqnarray}
Our predicted mass and decay properties of $B_s(1^3D_3)$ are
compatible with those predicted in Refs.~\cite{Xiao:2014ura,Lu:2016bbk,Asghar:2018tha}.
It should be pointed out that the large branching fraction for $Br[B_{sJ}(6064)\to B^*K]\simeq45\%$
seems to be not consistent with the observations naturally. Since $B_{sJ}(6064)$ causes
a clear bump structure around $6064$ MeV in the $B^+K^-$ mass spectrum through
the $B^+K^-$ decay, it should also cause another narrow bump structure
around $6019$ MeV through the $B^{*+}K^-$ decay with a missing photon from
$B^{*+}\to B^+\gamma$, however, this structure was not observed at LHCb. Thus,
it indicates that the $B_s(1^3D_3)$ may not be the main contributor to the $B_{sJ}(6064)$ structure
observed in the $B^+K^-$ mass spectrum.

As a whole, the $B_J(5970)$ may be assigned as the $1^3D_3$ assignment, which can be tested by the partial
width ratio between $B\pi$ and $B^*\pi$. It may be a flavor partner of the $D_3^*(2750)$ and
$D_{s3}(2860)$ resonances listed in RPP~\cite{Zyla:2020zbs}.
Their masses might be systematically overestimated by a value of $\sim 100$ MeV in some quark models~\cite{Ebert:2009ua,Godfrey:2016nwn,Godfrey:2015dva,Godfrey:1985xj}.
There still exists a puzzle to identify the $B_{sJ}(6064)$ structure as the $B_s(1^3D_3)$ state,
although both the predicted mass and width seem to be consistent with the observations.
To establish the narrow $B_s(1^3D_3)$ state finally, the observations of both the $B^+K^-$ and
$B^{*+}K^-$ decays and their partial width ratio are crucial in future experiments.

\subsubsection{$1^3D_1$ states}

In the $B$ meson sector, the mass for the $B(1^3D_1)$ state is predicted to be $M=6056$ MeV
in our potential model calculations, which is comparable with the predictions
in Refs.~\cite{Kher:2017mky,Lu:2016bbk,Asghar:2018tha,DiPierro:2001dwf}.
Our predicted mass for $B(1^3D_1)$ is about $80$ MeV larger than that for $B(1^3D_3)$.
Taking the mass $M=6056$ MeV, we calculate the
strong and radiative decay properties of $B(1^3D_1)$, our results are listed
in Table~\ref{decay2d1}. It is found that the $B(1^3D_1)$ state is a broad state with
a width of $\Gamma \simeq 350$ MeV, which is consistent
with the predictions in Refs.~\cite{Sun:2014wea,Lu:2016bbk}. This state mainly decays
into the $B\pi$, $B^*\pi$, and $B_1(5721)\pi$ channels with branching fractions
$\sim 26\%$, $12\%$, and $42\%$, respectively. The partial width ratio
between the two typical channels $B\pi$ and $B^*\pi$ is predicted to be
\begin{eqnarray}\label{ra}
\frac{\Gamma[B\pi]}{\Gamma[B^*\pi]}\simeq 2.2,
\end{eqnarray}
which may be helpful to identify the $B(1^3D_1)$ state from future observations.
In Refs.~\cite{Godfrey:2019cmi,Asghar:2018tha}, $B_J(5970)$ was suggested to
be a candidate for $B(1^3D_1)$. With this assignment
we find that the theoretical width $\Gamma\simeq 230$ MeV is too broad to
be comparable with the measured value  measured value $\Gamma_{exp}\simeq60-70$ MeV (see Table~\ref{Excitation}).
Thus, $B_{J}(5970)$ may not be a good candidate of the $1^3D_1$ state.

In the $B_s$ meson sector, the mass for the $B_s(1^3D_1)$ state is predicted to be $M=6101$ MeV
in our potential model calculations, which is comparable with the predictions
in Refs.~\cite{Kher:2017mky,Lu:2016bbk,DiPierro:2001dwf}.
Our predicted mass for $B_s(1^3D_1)$ is about $30$ MeV larger than that of $B_s(1^3D_3)$.
There are large uncertainties in the predictions of the mass splitting between
$B_s(1^3D_1)$ and $B_s(1^3D_3)$. In some works~\cite{Kher:2017mky,Zeng:1994vj,Asghar:2018tha},
the $B_s(1^3D_1)$ mass is even predicted to be smaller than that of $B_s(1^3D_3)$.
Taking the mass $M=6101$ MeV, we calculate the
strong and radiative decay properties of $B_s(1^3D_1)$, our results are listed
in Table~\ref{decay2d1}. It is found that the $B_s(1^3D_1)$ state has
a width of $\Gamma \simeq 130$ MeV, and mainly decays into the $BK$
and $B^*K$ channels with branching fractions $\sim 65\%$ and $27\%$, respectively.
The partial width ratio
between the two typical channels $BK$ and $B^*K$ is predicted to be
\begin{eqnarray}\label{ra}
\frac{\Gamma[BK]}{\Gamma[B^*K]}\simeq 2.4,
\end{eqnarray}
which is comparable with the predictions in Refs.~\cite{Sun:2014wea,Lu:2016bbk}.
From the point of view of mass, the observed resonance $B_{sJ}(6114)$ by the
LHCb Collaboration~\cite{Aaij:2020hcw} is a good candidate for the
$B_s(1^3D_1)$ state. While, the theoretical width $\Gamma \simeq 130$ MeV is also close
the upper limit of the measured width $\Gamma_{exp}=(66\pm 39)$ MeV.
There may exist a configuration mixing between the $2^3S_1$ and $1^3D_1$ states,
which will be discussed in the last part of this section.

\subsubsection{$1^1D_2$-$1^3D_2$ mixing}

There is a strong configuration mixing between the $1^3D_2$ and $1^1D_2$ states
for the heavy-light mesons predicted in the potential models. For the $B$ meson sector,
with the mixing scheme defined in Eq.~(\ref{mixst}) the mixing angle
is predicted to be $\theta_{1D}=-39.5^{\circ}$, which is close to the value $-50.8^{\circ}$
extracted in the heavy quark symmetry limit~\cite{Matsuki:2010zy,Close:2005se,Sun:2014wea}.
Our predicted masses for the $B(1D_2)$ and $B(1D'_2)$ states are about 5973
and 6067 MeV, respectively, which are close to the predictions in
Refs.~\cite{Kher:2017mky,DiPierro:2001dwf}. A fairly large mass splitting between
$B(1D_2)$ and $B(1D'_2)$, $\Delta M=94$ MeV, is obtained in present work. It
is comparable with the predictions of $\Delta M=110-130$ MeV in Refs.~\cite{Godfrey:2016nwn,Lu:2016bbk}.
With the masses and wave functions obtained from our potential model calculations,
the decay properties for these two mixed states $B(1D_2)$ and $B(1D'_2)$ are estimated,
the results are listed in Table~\ref{decay2d1}. It is found that the low
mass state $B(1D_2)$ has a broad width of $\Gamma\simeq 170$ MeV, and dominantly
decays into $B^*\pi$ and $B^*_2(5747)\pi$ channels with branching fractions about $54\%$
and $38\%$, respectively. While the high mass state $B(1D'_2)$ has a relatively narrow
width of $\Gamma\simeq 90$ MeV, and dominantly decays into $B^*\pi$, $B_1(5721)\pi$ and
$B^*_2(5747)\pi$ channels with branching fractions about $73\%$, $11\%$
and $7\%$, respectively. Our predicted decay properties are
roughly comparable with those predictions in Refs.~\cite{Xiao:2014ura,Sun:2014wea,Godfrey:2016nwn}.

For the $B_s$ meson sector, we predict the mixing angle
$\theta_{1D}= -39.9^{\circ}$. Our predicted masses for the $B_s(1D_2)$ and $B_s(1D'_2)$ states are about 6061
and 6113 MeV, respectively, which are close to the predictions in
Refs.~\cite{Lu:2016bbk,Zeng:1994vj,DiPierro:2001dwf}. An intermediate mass splitting between
$B_s(1D_2)$ and $B_s(1D'_2)$, $\Delta M=52$ MeV, is obtained in present work, which
is comparable with the predictions of $\Delta M=45-70$ MeV in Refs.~\cite{Lu:2016bbk,DiPierro:2001dwf}.
With the masses and wave functions obtained from our potential model calculations,
the decay properties for these two mixed states $B_s(1D_2)$ and $B_s(1D'_2)$ are estimated,
the results are listed in Table~\ref{decay2d1}. It is found that the low mass state $B_s(1D_2)$(6061)
has a intermediate width of
\begin{eqnarray}\label{dmx1}
\Gamma\simeq 90 \ \ \mathrm{MeV},
\end{eqnarray}
and dominantly decays into $B^*K$ channel with a branching fraction about $94\%$.
Our predicted width is about a factor $1.5-2$ smaller than those
predictions in Refs.~\cite{Asghar:2018tha,Lu:2016bbk,Sun:2014wea,Godfrey:2016nwn}.
The $B_s(1D_2)$ state may be observed around $6016$ MeV in the $B^{+}K^-$ mass spectrum through the
$B^{*+}K^-$ decay with a missing photon from $B^{*+}\to B^+\gamma$.

While the high mass state $B_s(1D'_2)$(6113) has a very narrow width of
\begin{eqnarray}\label{dmx}
\Gamma\simeq 23 \ \ \mathrm{MeV},
\end{eqnarray}
and dominantly decays into $B^*K$ channel with a branching fraction about $95\%$.
The decay properties predicted in present work are
in good agreement with those predictions in Refs.~\cite{Asghar:2018tha,Sun:2014wea,Xiao:2014ura,Godfrey:2016nwn}.
The narrow mixed state $B_s(1D'_2)$ with a mass of $M=6113$ MeV may be the main
contributor to the $B_{sJ}(6064)$ structure observed in
the $B^{+}K^-$ mass spectrum at LHCb~\cite{Aaij:2020hcw}. In this case, the
signal in the $B^{+}K^-$ mass spectrum may mainly come from the $B^{*+}K^-$
decay with a missing photon from $B^{*+}\to B^+\gamma$. Including the
energy of the missing photon, the mass and width are determined to be
$M_{exp}=(6109\pm 1.8)$ MeV and $\Gamma_{exp}=(22\pm 9)$ MeV for the resonance
$B_{sJ}(6109)$~\cite{Aaij:2020hcw}. It is interestingly found that
the $B_{sJ}(6109)$ resonance favors the assignment of the $1D$-wave
mixed state $B_s(1D'_2)$. The predicted mass, width, decay mode are
in good agreement with the observations. Finally, it should be
mentioned that the $B_s(1^3D_3)$ state may have a few contributions to
the $B_{sJ}(6064)$ structure through the $B^+K^-$ decay as well,
since this state with a mass of $M\simeq6067$ MeV lies just around
the peak position.

\begin{figure}[htbp]
\begin{center}
\centering  \epsfxsize=8.5cm \epsfbox{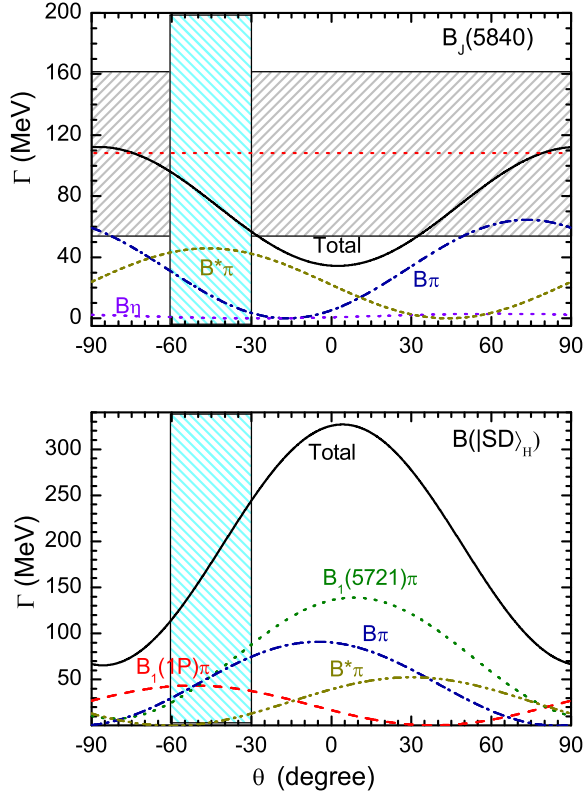}
\vspace{-0.2cm}\caption{The partial decay widths and total decay widths for the mixed states via $2^3S_1-1^3D_1$ mixing in the $B$-meson family as functions of the mixing angle $\theta$. In the horizontal direction, the shaded region represents the
possible range of the measured width from LHCb. In the vertical direction, shaded region represents the possible range of the mixing angle $\theta\simeq -(45\pm 16)^\circ$ suggested in Refs.~\cite{Zhong:2010vq,Zhong:2009sk}. The masses for $B_J(5840)$ and $B(|SD\rangle_H)$ are taken to be 5890 and 6040 MeV, respectively.} \label{5840}
\end{center}
\end{figure}

\begin{figure}[htbp]
\begin{center}
\centering  \epsfxsize=8.5cm \epsfbox{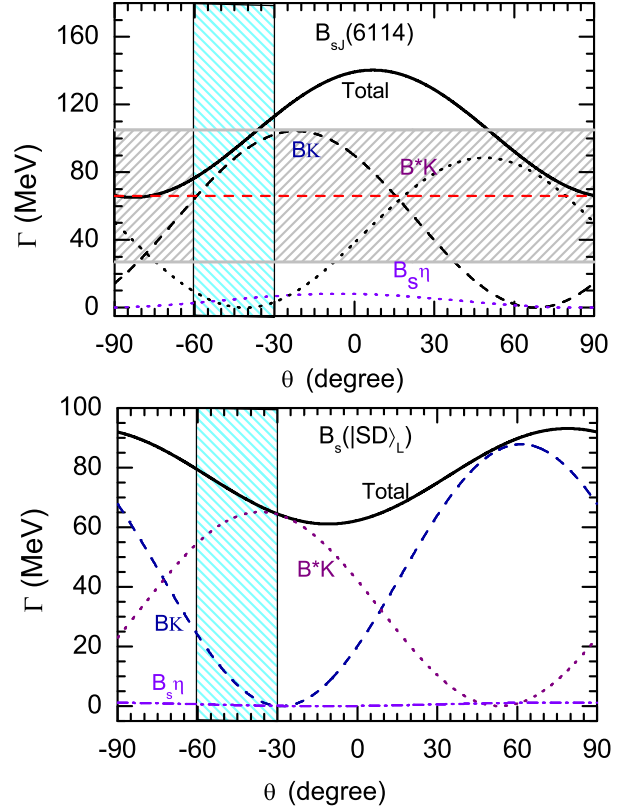}
\vspace{-0.2cm}\caption{The partial decay widths and total decay widths for the mixed states via $2^3S_1-1^3D_1$ mixing in the $B_s$-meson family as functions of the mixing angle $\theta$. In the horizontal direction, the shaded region represents the
possible range of the measured width from LHCb. In the vertical direction, shaded region represents the possible range of the mixing angle $\theta\simeq -(45\pm 16)^\circ$ suggested in Refs.~\cite{Zhong:2010vq,Zhong:2009sk}.  The masses for $B_{sJ}(6114)$ and $B_s(|SD\rangle_L)$ are taken to be 6114 and 5964 MeV, respectively.} \label{6114}
\end{center}
\end{figure}

\subsection{$2^3S_1$-$1^3D_1$ mixing}\label{sdmix}

It should be mentioned that there may exist a configuration mixing
between the $2^3S_1$ and $1^3D_1$ states. To explain the strong decay properties of the $D_J^*(2600)$ and/or
$D_{s1}(2700)$, configuration mixing between $2^3S_1$ and $1^3D_1$
is suggested in the literature~\cite{Close:2006gr,Chen:2011rr,Li:2009qu,Yu:2020khh,Chen:2015lpa}.
In Refs.~\cite{Zhong:2010vq,Zhong:2009sk}, our group
also carefully studied the strong decay properties of the $D_J^*(2600)$ and
$D_{s1}(2700)$. According the analysis, both $D_J^*(2600)$ and
$D_{s1}(2700)$ could be explained as the mixed state $|SD\rangle_L$ via the $2^3S_1$-$1^3D_1$
mixing with the following mixing scheme:
\begin{equation}\label{BsSD}
\left(
  \begin{array}{c}
  |SD\rangle_L\\
  |SD\rangle_H\\
  \end{array}\right)=
  \left(
  \begin{array}{cc}
   \cos\theta &\sin\theta\\
  -\sin\theta &\cos\theta\\
  \end{array}
\right)
\left(
  \begin{array}{c}
  2^3S_1\\
  1^3D_1\\
  \end{array}\right),
\end{equation}
where the mixed angle is estimated to be $\theta\simeq -(45\pm 16)^\circ$.
The $D_{s1}(2860)$ resonance observed in the $BK$ final state at LHCb~\cite{Aaij:2014xza,Aaij:2014baa}
seems to be the high mass mixed state $|SD\rangle_H$ as the partner of the low mass state $D_{s1}(2700)$
~\cite{Xiao:2014ura}, from which one can estimate a mass splitting $\Delta M\simeq 150$ MeV
between the high and low mass mixed states. Similarly, the $2^3S_1$-$1^3D_1$ mixing might also
exist in the $B$ and $B_s$ meson families.

The strong decay properties for the mixed states $|SD\rangle_L$
and $|SD\rangle_H$ in the $B$ and $B_s$ meson families were studied in another
work of our group~\cite{Xiao:2014ura}. It is interestingly found that the newly observed resonances $B_{J}(5840)$ and $B_{sJ}(6114)$
at LHCb~\cite{Aaij:2015qla,Aaij:2020hcw} are most likely to be the mixed states $B(|SD\rangle_L)$ and
$B_s(|SD\rangle_H)$, respectively, by comparing
the measured masses and widths with the theoretical predictions (see Figs. 1 and 2 in Ref.~\cite{Xiao:2014ura}).

Considering the $B_{J}(5840)$ resonance as the low mass mixed state $B(|SD\rangle_L)$,
we revise the strong decay properties by using the wave functions obtained
from our quark potential model calculations.
Our results are shown in Fig.~\ref{5840}. With the mixing angle
$\theta\simeq -(45\pm 16)^\circ$ determined in Refs.~\cite{Zhong:2010vq,Zhong:2009sk} and the
mass $M_{exp}\simeq 5890$ MeV measured at LHCb~\cite{Aaij:2015qla},
the $B(|SD\rangle_L)$ state has a width of $\Gamma\simeq (76\pm 20)$ MeV,
and dominantly decays into $B^*\pi$ channel. There may be a sizeable decay rate into
the $B\pi$ channel. The partial width ratio between $B\pi$ and $B^*\pi$ is predicted to be
\begin{eqnarray}\label{ra}
\frac{\Gamma[B\pi]}{\Gamma[B^*\pi]}\simeq 0.1-0.7,
\end{eqnarray}
which is sensitive to the mixing angle. The predicted width is consistent with the
measured width $\Gamma_{exp}=(107\pm 54)$ MeV by assuming $P=(-1)^J$. Moreover, the predicted decay modes are also consistent with
the observations of $B_{J}(5840)$. To better understand the nature of
$B_{J}(5840)$, more accurate measurements of the width together with the partial width ratio are expected to be carried out in
future experiments.

If $B_{J}(5840)$ corresponds to the low mass state $B(|SD\rangle_L)$ indeed, the mass of
$B(|SD\rangle_H)$ may be about 150 MeV larger than that of $B_{J}(5840)$.
Taking a mass of $M\simeq 6040$ MeV for $B(|SD\rangle_H)$,
we show its decay properties in Fig.~\ref{5840} as well. Within the mixing angle
range $\theta\simeq -(45\pm 16)^\circ$ suggested in~\cite{Zhong:2010vq,Zhong:2009sk},
the $B(|SD\rangle_H)$ has a width of $\Gamma\simeq (197\pm 47)$ MeV,
and mainly decay into $B\pi$ and $B_1(5721)\pi$ channels. The partial width ratio
between $B\pi$ and $B_1(5721)\pi$ is predicted to be
\begin{eqnarray}\label{ra}
\frac{\Gamma[B\pi]}{\Gamma[B_1(5721)\pi]}\sim 1,
\end{eqnarray}
which is insensitive to the mixing angle. Future observations in the $B\pi$ channel
with a larger data sample at LHCb may have a potential to discover this high mass mixed
state $B_{1}(|SD\rangle_H)$.

In the $B_s$ meson sector, considering the $B_{sJ}(6114)$ resonance observed
in the $B^+K^-$ mass spectrum~\cite{Aaij:2020hcw} as the high mass mixed state $B_s(|SD\rangle_H)$,
we revise the strong decay properties by using the wave functions obtained
from our quark potential model calculations.
Our results are shown in Fig.~\ref{6114}. It is found that with the mixing angle
$\theta\simeq -(45\pm 16)^\circ$ determined in Refs.~\cite{Zhong:2010vq,Zhong:2009sk},
the $B_s(|SD\rangle_H)$ state has a width of $\Gamma\simeq (95\pm 15)$ MeV,
and dominantly decays into $BK$ channel with a branching fraction $\sim90\%$.
Both the decay mode $B^+K^-$ and width $\Gamma_{exp}=(66\pm 39)$ MeV observed
for $B_{sJ}(6114)$ at LHCb~\cite{Aaij:2020hcw} can be well understood in our quark model calculations.
Thus, the $B_{sJ}(6114)$ may favor the mixed state $B_s(|SD\rangle_H)$.

The mass for the low mass state $B_{s1}(|SD\rangle_L)$ may be about 150 MeV smaller than
that of $B_{sJ}(6114)$. Taking a mass of $M\simeq5960$ MeV for the low mass state $B_{s1}(|SD\rangle_L)$,
we show its decay properties in Fig.~\ref{6114} as well. In the mixing angle range $\theta\simeq -(45\pm 16)^\circ$,
the $B_{s1}(|SD\rangle_L)$ has a width of $\Gamma\simeq (70\pm 10)$ MeV,
and mainly decay into $B^*K$ channel. The partial width ratio between $BK$ and $B^*K$,
\begin{eqnarray}\label{ra}
\frac{\Gamma[BK]}{\Gamma[B^*K]}< 0.5,
\end{eqnarray}
is sensitive to the mixing angle. The $B_{s1}(|SD\rangle_L)$ is most likely to be observed
in the $B^+K^-$ final state with a larger data sample at LHCb.

As a whole the $B_{J}(5840)$ and $B_{sJ}(6114)$ may favor the mixed
states $B(|SD\rangle_L)$ and $B_s(|SD\rangle_H)$ via $2^3S_1$-$1^3D_1$ mixing, respectively.
Their partners $B(|SD\rangle_H)$ and $B_s(|SD\rangle_L)$ are expected to
be observed in their dominant decay channels with a larger data sample at LHCb.

%In summary, the $B_{sJ}(6114)$ resonance may favor a mixed state .
%The lower $B_s(2^3S_1)-B_s(1^3D_1)$ mixing state $B_{sJ}(|SD\rangle)$ might have a a relatively narrow width and are most likely to be observed in future experiment. Our conclusion is in good agreement with our previous work~\cite{Xiao:2014ura}.

\section{summary}\label{SUM}

The experimental progress provides us good opportunities to establish
an abundant $B$ and $B_s$-meson spectrum up to the second orbital excitations.
In this work, combining the newest experimental progress, we carry out a systematical
study of the mass spectrum, strong decays and radiative decays of the
$1P$-, $1D$-, and $2S$-wave excited $B$ and $B_s$ states in the constitute quark model.
The mass and strong decay properties for the well established $1P$-wave resonances $B_1(5721)^{+,0}$, $B_2^*(5747)^{+,0}$,
$B_{s1}(5830)$ and $B_{s2}^*(5840)$ can be consistently explained.
The possible assignments for the high mass resonances/structures
$B_J(5840)^{0,+}$, $B(5970)^{0,+}$, $B_{sJ}(6064)$ and $B_{sJ}(6114)$
are discussed. We hope that our study can provide some useful information
towards establishing an abundant $B$ and $B_s$-meson spectrum.
Our main results are summarized as follows.

For the $P$-wave states, several points should be emphasized.
(i) Some radiative decay processes, such as $B_{s1}(5830)\to B_s^{(*)}\gamma$
and $B_{s2}^*(5840)\to B_s^*\gamma$, have good potentials
to be found in future experiments due to their fairly branching fractions of $\mathcal{O}(10^{-2})$.
(ii) The $B_{J}(5732)$ and $B_{sJ}(5850)$ resonances listed by the PDG~\cite{Zyla:2020zbs}
may good candidates for the missing $|1P_1\rangle$ state in the $B$ and $B_s$ families, respectively.
(iii) Both $B(1^3P_0)$ and $B_s(1^3P_0)$ may hardly be observed in experiments due
to their very broad width of $\Gamma\sim300$ MeV.

The $B_{J}(5840)$ resonance and the new $B_{sJ}(6114)$ structure
observed in the $B^+K^-$ mass spectrum may be explained with the mixed
states $B(|SD\rangle_L)$ and $B_s(|SD\rangle_H)$ via $2^3S_1$-$1^3D_1$ mixing, respectively.
To confirm the nature of the $B_{J}(5840)$ and $B_{sJ}(6114)$,
the typical ratios $\Gamma(B\pi)/\Gamma(B^*\pi)$ and $\Gamma(B^*K)/\Gamma(BK)$
are suggested to be measured in future experiments. The other two missing states, $B(|SD\rangle_H)$ with a
mass of $M\simeq 6010$ MeV and $B_s(|SD\rangle_L)$ with a mass of $M\simeq 5960$ MeV,
may be observed in the $B\pi$ and $B^*K$ final states. On the other hand, if there
is little mixing between $2^3S_1$-$1^3D_1$, the $B_{J}(5840)$ and $B_{sJ}(6114)$ resonances
may be candidates for the $B(2^3S_1)$ and $B_s(1^3D_1)$ states, respectively.

The $B_J(5970)$ resonance may be assigned as the $1^3D_3$ state in the $B$ meson family,
although it as a pure $2^3S_1$ state cannot be excluded according to present
experimental information. To clarify the nature of $B_J(5970)$, further observations
of the $B\pi$ and $B^*\pi$ channels and a measurement of their
partial width ratio are necessary. In the $B_s$ family, the predicted mass
$M\simeq6067$ MeV and width $\Gamma\simeq 13$ MeV for
the $B_s(1^3D_3)$ are consistent with the $B_{sJ}(6064)$ structure
observed in the $B^+K^-$ mass spectrum. However, the $B_s(1^3D_3)$ may not be
the main contributor to the $B_{sJ}(6064)$ structure. If $B_{sJ}(6064)$
corresponds to $B_s(1^3D_3)$, another narrow structure around $6019$ MeV coming from
the $B^{*+}K^-$ decay should be observed in the $B^+K^-$ mass spectrum, however, it was
not seen at LHCb.

The narrow $B_{sJ}(6064)$ structure observed in the $B^+K^-$ mass spectrum
may mainly come from the resonance $B_{sJ}(6109)$ decaying into $B^{*+}K^-$.
The $B_{sJ}(6109)$ resonance favors the assignment of the high mass $1D$-wave
mixed state $B_s(1D'_2)$ with $J^P=2^-$. This state dominantly decays into $B^*K$
channel with branching fraction about $95\%$, and the $BK$ decay is forbidden.
The other missing mixed state $B_s(1D_2)$ has a mass of $M\simeq5973$ MeV
and a width of $\Gamma\simeq 90$ MeV. It is most likely to be established in the $B^{+}K^-$ mass
spectrum through the $B^{*+}K^-$ decay with a missing photon from $B^{*+}\to B^+\gamma$.
In the $B$-meson sector, two relatively broad mixed states, $B(1D_2)$ with
$M\simeq5973$ MeV and $B(1D'_2)$ with $M\simeq 6067$ MeV, may be
observed their main decay channel $B^{*}\pi$ with a larger data sample at LHCb.

Finally, it should be mentioned that the $B_{J}(5840)$ resonance may be
a candidate of the $2S$-wave state $B(2^1S_0)$ as well.
In this case, the $B\pi$ mode of $B_{J}(5840)$ is forbidden, which should be
further confirmed in future experiments. In the $B_s$ meson sector, our predicted mass and width for the $B_s(2^1S_0)$ state
are $M=5944$ MeV and $\Gamma\simeq 55$ MeV, respectively.
The $B^*K$ channel is the only OZI-allowed two body strong decay mode.
The $B_s(2^1S_0)$ state should have large potentials to be seen in the $B^{*+}K^-$ channel
since it has a fairly narrow width.

%In theory, the mass and
%width of $B(2^1S_0)$ are predicted to be $M=5876$ MeV and $\Gamma\simeq 63$ MeV, respectively, which are
%consistent with observations of $B_{J}(5840)$.

\section*{Acknowledgement}

Helpful discussions with Qi-Fang L\"{u} and Ming-Sheng Liu are greatly appreciated.
This work is supported by the National Natural Science Foundation of China (Grants Nos. U1832173, 11775078).

\end{document}